\documentclass[]
{jpp}

\usepackage{graphicx}
\usepackage{epstopdf, epsfig}
\usepackage[utf8]{inputenc}
\usepackage[T1]{fontenc}
\usepackage{amsmath}
\usepackage[usenames,dvipsnames]{color}
\usepackage{float}
\usepackage{mathtools}
\usepackage{amsmath,amssymb,amsbsy}
\usepackage{bbm,bm,mathrsfs,yfonts}
\usepackage[capitalize]{cleveref}
\usepackage{xcolor}
\usepackage{microtype}
\usepackage[normalem]{ulem}
\newcommand\redsout{\bgroup\markoverwith{\textcolor{red}{\rule[0.5ex]{2pt}{0.4pt}}}\ULon}

\newcommand{\be}{\begin{equation}}
\newcommand{\ee}{\end{equation}}

\renewcommand{\l }{\left}
\renewcommand{\r }{\right}

\DeclarePairedDelimiter\floor{\lfloor}{\rfloor}

\shorttitle{Collisional Damping of Linear EPW}
\shortauthor{R. Jorge et al.}

\title{Linear Theory of Electron-Plasma Waves at Arbitrary Collisionality}

\author{
  R. Jorge\aff{1,2}\corresp{\email{rogerio.jorge@epfl.ch}}\thanks{Now present at Institute for Research in Electronics and Applied Physics, University of Maryland, College Park MD 20742, USA},
  P. Ricci\aff{1},
  S. Brunner\aff{1},
  S. Gamba\aff{1,3},
  V. Konovets\aff{1},
  N. F. Loureiro\aff{4},
  L. M. Perrone\aff{1},
  \and N. Teixeira\aff{2}
}

\affiliation{
\aff{1}École Polytechnique Fédérale de Lausanne (EPFL), Swiss Plasma Center (SPC), CH-1015
Lausanne, Switzerland
\aff{2}Instituto de Plasmas e Fusão Nuclear, Instituto Superior Técnico, Universidade de Lisboa, 1049-001 Lisboa, Portugal
\aff{3}Department of Energy, Politecnico di Milano, Via Ponzio 34/3, Milano, 20133, Italy
\aff{4}Plasma Science and Fusion Center, Massachusetts Institute of Technology, Cambridge MA 02139, USA
}

\begin{document}

\maketitle

\begin{abstract}
The dynamics of electron-plasma waves {are} described at arbitrary collisionality by considering the full Coulomb collision operator.
The description is based on a Hermite-Laguerre decomposition of the velocity dependence of the electron distribution function.
The damping rate, frequency, and eigenmode spectrum {of} electron-plasma waves are found as function{s} of the collision frequency and wavelength.
A comparison is made {between} the collisionless Landau damping limit, the Lenard-Bernstein and Dougherty collision operators, and the electron-ion collision operator, finding large deviations in the damping rates and eigenmode spectra.
A purely damped entropy mode, characteristic of a plasma where pitch-angle scattering effects are dominant {with respect to collisionless effects}, is shown to emerge numerically, and its dispersion relation is analytically derived.
It is shown that such {a} mode is absent when simplified collision operators are used, and that like-particle collisions strongly influence the damping rate of the entropy mode.
\end{abstract}

\section{Introduction}

Electron-plasma waves (EPW), also called Langmuir waves or plasma oscillations, are oscillations of the electron density at the plasma frequency resulting from the break of local charge neutrality \citep{Bohm1949,Malmberg1966}.
The displacement of electrons leads to an electrostatic force that, by pulling electrons back to their equilibrium position, results into oscillations of the electrostatic potential and electron density.
In a collisionless system, the amplitude of EPW decreases with time due to Landau damping \citep{Landau1946}.
The phenomenon of collisionless Landau damping  is well understood, both linearly and non-linearly \citep{Dawson1961,ONeil1965,Zakharov1972,Morales1972,Mouhot2011a}.
When Coulomb collisions are present, {although} collisional and Landau damping of EPW {are known to} act synergistically {\citep{Brantov2012}, the physical mechanisms which dictate their interplay are considerably less understood}.
This is despite the fact that
{plasmas are all characterized by a finite number of particles in a Debye sphere, $N=(4\pi/3)n \lambda_D^3$, with $n$ the electron density and $\lambda_D$ the Debye length, and therefore collisional effects are always present.
Indeed,}
understanding the behavior of EPW with collisions is important since Coulomb collisions significantly contribute to the behavior of many important laboratory plasmas, such as magnetic \citep{Scott2007} and inertial fusion \citep{Lindl2004} plasmas, {and} plasmas for industrial processing \citep{Lieberman2005}.
Collisions also influence the dynamics of EPW in near-earth space plasmas \citep{Jordanova1996}, and can even be the only source of significant damping {of EPW in low-temperature laboratory plasmas} \citep{Banks2017}.

The need for a simplified theoretical framework able to describe Coulomb collisions at arbitrary collisionality is widely recognized, and has been the subject of considerable interest over the past few decades \citep{Callen1997, Ji2010}, with a large effort devoted not only to the study of EPW \citep{Hammett1990,Brantov2012,Banks2016}, but also to ion-acoustic waves \citep{Epperlein1992,Tracy1993a,Zheng2000}, and drift-waves \citep{Jorge2018}.
The difficulty associated with an accurate estimate of the collisional damping in a plasma at arbitrary collisionalities is related to the integro-differential character of the Coulomb collision operator $C(f)$ \citep{Helander2002}.
While progress can be made using fluid models that rely on the evaluation of the velocity moments of the kinetic equation \citep{Braginskii1965}, {the standard formulation of these models based on collisional closures} assumes that {typical wave-numbers $k$ of the system are small compared with the inverse mean-free path $1/\lambda_{mfpa}=\nu_{a}/v_{tha}$, with $\nu_a$ the collision frequency and $v_{tha}$ the thermal velocity of the species $a$, and that typical frequencies $\omega$ are small compared with $\nu_a$.}
%
This {restricts} the application of {standard} fluid models to highly collisional regimes, therefore excluding Landau damping effects.
In order to incorporate kinetic effects in fluid models, closures that mimic the linear response of a collisionless plasma have been derived \citep{Hammett1992a,Hammett1993}, later extended to include fourth order moments \citep{Hunana2018}, and to include collisional effects without pitch-angle scattering \citep{Joseph2016}.

{A possible approach to the study of the kinetic properties of EPW is based on the development of the distribution function on a {convenient} basis, and the projection of the kinetic equation on this basis.
Indeed,}
pseudospectral decompositions that expand the electron distribution function in an appropriate orthogonal polynomial basis {have allowed a rigorous assessment of} the effect of {collisional} pitch-angle scattering in {linear} EPW and {ion-acoustic waves} by including electron-ion collisions while neglecting electron-electron collisions (justified in a high-Z regime) \citep{Epperlein1992,Banks2016}.
The role of self-collisions {in the linear regime} was investigated in \citet{Banks2017} using a simplified operator with respect to the full Coulomb operator and in \citet{Brantov2012} where a simplified form for the high order moments of the like-species Coulomb collision operator was employed in order to derive an analytic dispersion relation.

In this work, the {linear} properties of EPW are assessed by using the full {linearized} Coulomb electron-electron and electron-ion collision operators at arbitrary collision frequencies.
For this purpose, a pseudospectral decomposition of the electron distribution function based on a Hermite-Laguerre polynomial basis is used.
{Leveraging the work in \citet{Jorge2017,Jorge2018}}, a moment expansion of the full Coulomb collision operator is performed at all orders by taking into account both like-species and inter-particle collisions without simplifying assumptions.
The framework used here allows, {for the first time}, the evaluation of the frequency and damping rates and, more {generally}, of the linear spectrum, of EPW eigenmodes, {at arbitrary collisionalities}.
Among the subdominant modes, we focus on the analytical and numerical description of the entropy mode, a purely damped mode that requires the Coulomb collision operator to be properly described \citep{Epperlein1994,Banks2016}.
{The entropy mode can have a damping rate comparable to other modes in the plasma [such as ion-acoustic waves \citep{Tracy1993a}] and similar wave-numbers, and it determines the damping rate of the system on collisional time scales.}
In fact, as we show, this mode is absent when the kinetic equation is solved using approximate collision operators or in one-dimensional velocity space descriptions and, in general, deviations between the results based on the Coulomb and simplified collision operators (such as the Lenard-Bernstein, the Dougherty, and the electron-ion operators) are particularly evident.
We remark that the discrepancies in the spectrum observed between {different} collision operators may lead to major differences in the nonlinear evolution of EPW.
Indeed, stable modes can be non-linearly excited to a finite amplitude and have a major role in nonlinear energy dissipation and turbulence saturation, affecting the formation of turbulent structures, as well as heat and particle transport \citep{Terry2006,Hatch2011}.
As a test of our numerical investigations, the results for the Lenard-Bernstein case are compared to the eigenmode spectrum resulting from an analytical solution where the plasma distribution function and the electrostatic potential are decoupled.
This also allows us to gain some insight on previous EPW results using the Lenard-Bernstein operator \citep{Bratanov2013,Schekochihin2016}.
In addition, we compare our pseudospectral decomposition to the one based on a Legendre polynomial expansion for the case of the electron-ion operator.

This paper is organized as follows.
\cref{sec:momhierarchy} presents the moment-hierarchy equation used for the EPW description, deriving it from the kinetic Boltzmann equation by using a Hermite-Laguerre expansion of the electron distribution function.
\cref{sec:collisionless} focuses on the collisionless moment-hierarchy and derives the collisionless dispersion relation.
In \cref{sec:simresults}, the oscillation frequency and damping rates of EPW are analyzed and compared with simplified collision operators.
\cref{sec:timeevol} derives a dispersion relation for the entropy mode that shows remarkable agreement with the numerical results.
Finally, \cref{sec:eigspec} shows the EPW eigenvalue spectrum using different collision operators and discretization methods.
The conclusions follow.

\section{Moment-Hierarchy Formulation of EPW}
\label{sec:momhierarchy}

We briefly describe the Boltzmann-Poisson system, our starting point for the description of EPW, for an unmagnetized plasma, and derive a moment expansion of the distribution function that allows its {numerical} solution.
The Boltzmann equation for the evolution of the electron distribution function $f$ is given by
\begin{equation}
    \frac{\partial f}{\partial t} + \mathbf v \cdot \nabla f + \frac{e}{m}\nabla \phi \cdot \frac{\partial f}{\partial \mathbf v} = \hat C(f).
\label{eq:boltzmann}
\end{equation}
In \cref{eq:boltzmann}, $e$ is the elementary charge, $m$ the electron mass, $\phi$ the electrostatic potential, and $\hat  C(f)$ the non-linear Coulomb {(also called Landau)} collision operator for electrons
\begin{equation}
    \begin{split}
        \hat C(f)={\frac{v_{th}^3}{n}}\sum_b{\nu_{b}}&\partial_{\mathbf v} \cdot \l[\frac{m}{m_b}({\partial_{\mathbf v} H_b})f - \partial_{\mathbf v}(\partial_{\mathbf v}G_b)\cdot \partial_{\mathbf v}f\r],
    \end{split}
    \label{eq:caa}
\end{equation}
{where $v_{th}=\sqrt{2T/m}$ is the electron thermal velocity with $T$ the electron temperature, and} $\nu_{b}$ the characteristic collision frequency between electrons and species $b$ ($b=e,i$ for electrons and ions, respectively), defined by
{
\begin{equation}
    \nu_e=\frac{8\sqrt{\pi}}{3 \sqrt{m_e}}\frac{n \lambda e^4}{T^{3/2}},
\end{equation}
with $\lambda$ the Coulomb logarithm and $\nu_i=\sqrt{2}\nu_e$.
}
$H_b=2\int {f_b(\mathbf v')}/{|\mathbf v - \mathbf v'|}d\mathbf v'$  and $G_b=\int f_b(\mathbf v')|\mathbf v - \mathbf v'|d\mathbf v'$ are the Rosenbluth potentials \citep{Helander2002}.
We relate the electrostatic potential $\phi$ to $f$ using Poisson's equation
\begin{equation}
    \nabla^2 \phi = 4 \pi e \left(\int f d \bm v - n_0\right),
\end{equation}
where the ions are assumed to provide a fixed homogeneous, neutralizing background, with density $n_0$ and a Maxwellian-Boltzmann equilibrium with the same temperature as the electrons.
{An atomic number $Z=1$ is considered.}
Equation (\ref{eq:boltzmann}) is linearized by expressing $f$ as $f=f_M(1+\delta f)$ with $\delta f \ll 1$ and $f_M$ an isotropic Maxwell-Boltzmann equilibrium distribution with constant density $n_0$ and temperature $T_0$, yielding
\begin{equation}
    f_M\frac{\partial \delta f}{\partial t} + f_M \mathbf v \cdot \nabla \delta f + \frac{e}{m}\nabla \delta \phi \cdot \frac{\partial f_M}{\partial \mathbf v} = C(f_M \delta f),
\label{eq:linboltzmann}
\end{equation}
where we used the fact that $\hat C(f_M)=0${, and $C(f_M \delta f)$ is the linearized version of the Coulomb collision operator in \cref{eq:caa} whose expression can be found in \citet{Helander2002}}.
The Boltzmann equation, \cref{eq:linboltzmann}, is coupled to the Poisson equation $\nabla^2 \delta \phi= 4\pi e \delta n$ with $\delta n=\int f_M \delta f d \mathbf v$ the perturbed electron density.
We now rewrite \cref{eq:linboltzmann} in terms of the Fourier transformed distribution function $\delta f_k=\int \delta f \exp({-i \mathbf k \cdot \mathbf x}) d\mathbf{x}$ as
\begin{equation}
    \frac{\partial \delta f_k}{\partial t} + i \mathbf k \cdot \mathbf v \delta f_k+i \mathbf k \cdot \mathbf v \frac{4 \pi e^2 \delta n_k}{k^2 T_0} = \frac{C(f_M \delta f_k)}{f_M},
\label{eq:fourgyrolinboltzmann}
\end{equation}
where we used the Fourier transformed Poisson equation $-k^2 \delta \phi_k = 4 \pi e \delta n_k$, with $\delta n_k=\int \delta n \exp({-i \mathbf k \cdot \mathbf x}) d\mathbf{x}$ and $\delta \phi_k=\int \delta \phi \exp({-i \mathbf k \cdot \mathbf x}) d\mathbf{x}$.

Similarly to previous studies on the collisional damping of EPW \citep{Brantov2012,Banks2016}, a three-dimensional cylindrical $(v_\perp, \varphi, v_z)$ velocity coordinate system is used, therefore decomposing the velocity vector $\mathbf v$ as
\begin{equation}
    \mathbf v = v_z \mathbf e_z + v_\perp(\cos \varphi \mathbf e_x + \sin \varphi \mathbf e_y),
\label{eq:vdecomp}
\end{equation}
{where} $(\mathbf e_x, \mathbf e_y, \mathbf e_z)$ {are} {Cartesian} unit vectors with $z$ the direction of the wave-vector $\bm k = k \bm e_z$.
{
In this work, we consider the azymuthally isotropic, i.e. $\varphi$ independent, subset of solutions $\delta f_k = \delta f_k(v_\perp,v_z)$ of the Boltzmann equation, \cref{eq:fourgyrolinboltzmann}, which further allows us to} reduce {our study to the one of} a two-dimensional model.
{This can be done by applying} the averaging operator $\left<...\right>$ defined by
\begin{equation}
    \left< g\right>(v_\perp,v_z) = \frac{1}{2\pi}\int_0^{2\pi} g(v_\perp,\varphi,v_z) d\varphi,
\label{eq:gyrooperator}
\end{equation}
to the Boltzmann equation, \cref{eq:linboltzmann}, yielding
\begin{equation}
    \frac{\partial \left<\delta f_k\right>}{\partial t} + i k v_z \left<\delta f_k\right> + i k v_z \frac{4 \pi e^2 \delta n_k}{k^2 T_0} = \frac{\left<C(f_M \delta f_k)\right>}{f_M}.
\label{eq:gyrolinboltzmann}
\end{equation}
{As the linearized Coulomb collision operator satisfies $\left<C(f_M \delta f_k)\right>=C(f_M \left<\delta f_k\right>)$, \cref{eq:gyrolinboltzmann} can be used to obtain the subset of azimuthally symmetric solutions in velocity space $\left<\delta f_k \right>$, which are decoupled from the azimuthally asymmetric solutions ${\delta \tilde f}_k = \delta f_k - \left< \delta f_k \right>$.}
Finally, we rewrite \cref{eq:gyrolinboltzmann} by normalizing time to $k v_{th0}$ with $v_{th0}=\sqrt{2 T_0/m}$ the {electron} thermal velocity, $\delta n_k$ to $n_0$, and $v_z$ to $v_{th}$, yielding
\begin{equation}
    i\frac{\partial \left<\delta f_k\right>}{\partial t} - v_z \left<\delta f_k\right>-\frac{ v_z{ \delta n_k}}{\alpha_D} = i\frac{ \left<C(f_M \delta f_k)\right>}{f_M},
\label{eq:boltzmanneq}
\end{equation}
where we defined $\alpha_D = k^2 \lambda_D^2$ with $\lambda_D=\sqrt{T_0/(4 \pi e^2 n_0)}$ the Debye length{, and where the collision frequency coefficient present in $C(f_M \delta f_k)$ is now in units of $k v_{th}$}.
{Furthermore, we define $\nu$ as the electron-ion collision frequency normalized to $k v_{th}$, namely
\begin{equation}
    \nu = \frac{\nu_i}{k v_{th}}.
\end{equation}
We note that the number of particles in a Debye-sphere $N$ can be cast in terms of $\nu$ and $\alpha_D$ as
\begin{equation}
    N = \frac{1}{9}\sqrt{\frac{2}{\pi}}\frac{\lambda}{\nu \sqrt{\alpha_D}},
\end{equation}
where $\lambda$ is the Coulomb logarithm.
}

{Following \citet{Jorge2017,Jorge2018}}, we solve the linearized kinetic equation, \cref{eq:boltzmanneq}, at arbitrary collisionalities by expanding the perturbed distribution function $\left<\delta f_k\right>$ into an orthogonal Hermite-Laguerre polynomial basis of the form
\begin{equation}
    \left<\delta f_k\right> = \sum_{p,j=0}^{\infty}\frac{N^{pj}}{\sqrt{2^p p!}}H_p\left(v_z\right)L_j\left(v_\perp^2\right),
\label{eq:gyrof}
\end{equation}
where $H_p$ are {\textit{physicists'}} Hermite polynomials of order $p$, defined by the {Rodrigues' formula}
\begin{equation}
    H_p(x)=(-1)^p e^{x^2}\frac{d^p}{dx^p}e^{-x^2},
\end{equation}
and normalized via {the orthogonality condition based on the scalar product with weight factor $e^{-x^2}$}
\begin{equation}
    \int_{-\infty}^{\infty} dx H_p(x) H_{p'}(x) e^{-x^2} = 2^p p! \sqrt{\pi} \delta_{p{p'}},
\end{equation}
and  $L_j$ the Laguerre polynomials of order $j$, defined by {the corresponding Rodrigues' formula}
\begin{equation}
    L_j(x)=\frac{e^x}{j!}\frac{d^j}{dx^j}(e^{-x}x^j),
\end{equation}
and orthonormal with respect to the weight $e^{-x}$
\begin{equation}
    \int_{0}^{\infty} dx L_j(x) L_{j'}(x) e^{-x} = \delta_{jj'}.
\end{equation}
Due to the orthogonality of the Hermite-Laguerre basis, the coefficients {$N^{pj}=N^{pj}(k,t)$} of the expansion in \cref{eq:gyrof} can be computed via the expression
\begin{equation}
    N^{pj}=\int \frac{H_p(v_z) L_j(v_\perp^2) \left<\delta f_k\right> }{\sqrt{2^p p!}} \frac{e^{-v_z^2-v_\perp^2}}{\sqrt{\pi}} dv_z dv_\perp^2.
    \label{eq:gyromoments}
\end{equation}
{With respect to a grid treatment using collocation points based on orthogonal polynomials \citep{Belli2008,Landreman2013}, we note that the Hermite-Laguerre decomposition used in this work allows us to find the Rosenbluth potentials analytically as linear combinations of the $N^{pj}$ moments of the distribution function, to find linear and nonlinear closures at an arbitrary number of moments and mean-free path, and to make much more evident the role of the parallel and perpendicular phase-mixing processes.}

By projecting the Boltzmann equation, \cref{eq:boltzmanneq}, onto a Hermite-Laguerre basis, a system of differential equations (henceforth called moment-hierarchy) for the coefficients $N^{pj}$ is obtained
\begin{align}
    i \frac{\partial}{\partial t} N^{pj} &= \sqrt{\frac{p+1}{2}}N^{p+1 j}+\sqrt{\frac{p}{2}}N^{p-1 j}+ \frac{N^{00}}{\alpha_D}\frac{\delta_{p,1}\delta_{j,0}}{\sqrt{2}}+i C^{pj},
\label{eq:momenthierarchy}
\end{align}
with $C^{pj}$ the projection of the linearized collision operator onto a Hermite-Laguerre basis
\begin{equation}
    C^{pj}=\int \frac{H_p(v_z) L_j(v_\perp^2) \left<C(f_M \delta f_k)\right> }{\sqrt{2^p p!}} dv_z dv_\perp^2.
\label{eq:projectioncoll}
\end{equation}

The Coulomb collisional moments $C^{pj}$ are derived leveraging the result in \citet{Ji2006}, where the collision operator is projected onto a tensorial basis of the form $\mathbf p^{ls}=\mathbf P^{l}(\mathbf c)L_s^{l+1/2}(c^2)$, with the {irreducible} tensorial Hermite polynomials defined by the recurrence $\mathbf P^{l+1}(\mathbf c)=\mathbf c \mathbf P^l(\mathbf c)-c^2 \partial_{\mathbf c}\mathbf P^l(\mathbf c)/(2l+1)$, being $\mathbf P^0(\mathbf c)=1$ with $\mathbf c = \mathbf v/v_{th}$, and the associated Laguerre polynomials $L_s^{l+1/2}(x)=\sum_{m=0}^s L_{sm}^l x^m$ with $L_{sm}^{l}=[{(-1)^m(l+s+1/2)!}]/[{(s-m)!(l+m+1/2)!m!}]$.
Indeed, expanding the distribution function as $\delta f_a=\sum_{l,s}\mathbf M_a^{ls} \cdot \mathbf p^{ls}/\sigma_s^l$, and with $\sigma_s^l=l!(l+s+1/2)!/(2^l(l+1/2)!s!)$, Ji and Held showed that the {linearized} collision operator can be written as
\begin{equation}
    C(f_M \delta f)=f_M \sum_b \sum_{l,s=0}^\infty\frac{\mathbf P^{l}(\hat v)}{\sigma_{s}^l}\cdot \left(\mathbf M_e^{ls} \nu_{eb}^{ls,0}+\mathbf M_b^{ls} \nu_{eb}^{0,ls}\right).
    \label{eq:jimoment}
\end{equation}
where $\nu_{eb}^{ls,0}(v)$ and $\nu_{eb}^{0,ls}(v)$ are linear combinations of the error function and its derivatives [for their expression, see \citet{Ji2006}]{, and represent the test-particle and field-particle (back-reaction) parts of the linearized collision operator, respectively}.
We remark that a similar expansion in Legendre-Associated Laguerre polynomials was used in \citet{Brantov2012} in order to derive a simplified dispersion relation applicable to the study of EPW, ion-acoustic waves, and entropy modes.

In order to evaluate $C^{pj}$, we Fourier transform in space and average the operator $C(f_M \delta f)$ in \cref{eq:jimoment} according to \cref{eq:gyrooperator}, using the averaging identity $\left< \mathbf P^l(\mathbf c) \right> = c^l P_l(v_z/v)\mathbf P^l(\hat e_z)$, with $P_l(x)=\partial^l_x(x^2-1)^l/(2^l l!)$ the Legendre polynomials.
This yields
\begin{equation}
    \left<C(f_M \delta f_k)\right> = \sum_{b}\sum_{l,s=0}^{\infty}(C_{eb}^{ls,0} + C_{eb}^{0,ls}),
\label{eq:jiheldfmf1}
\end{equation}
where $C_{eb}^{ls,0}$ is the averaged test-particle operator
\begin{equation}
    C_{eb}^{ls,0} = f_{M} \nu_{eb}^{ls,0}(v) P_l\left(\frac{v_z}{v}\right) M_e^{ls} \frac{2^l (l!)^2}{(2l!)},
\label{eq:jiheldtestp}
\end{equation}
and $C_{eb}^{0,ls}$ the field-particle {(back-reaction)} operator
\begin{equation}
    C_{eb}^{0,ls} = f_{M} \nu_{eb}^{0,ls}(v) P_l\left(\frac{v_z}{v}\right) M_b^{ls} \frac{2^l (l!)^2}{(2l!)},
\label{eq:jiheldfieldp}
\end{equation}
with the fluid moments defined as
\begin{equation}
    M_b^{ls} = \sum_{p=0}^{l+2s}\sum_{j=0}^{s+l/2}T_{ls}^{pj} N_b^{pj} \sqrt{\frac{2^p p!}{\sigma_s^l}}.
\label{eq:fluidmoments}
\end{equation}
The basis transformation coefficients $T_{ls}^{pj}$ in \cref{eq:fluidmoments} are defined by
\begin{equation}
    v^l P_l\left(\frac{v_z}{v}\right) L_s^{l+1/2}(v^2) =\sum_{p=0}^{l+2s}\sum_{j=0}^{s+l/2} T_{ls}^{pj} H_p(v_z) L_j(v_\perp^2),
\label{eq:basistransf}
\end{equation}
and their closed form expression can be found in \citet{Jorge2017}.
{While an expansion in tensorial Hermite polynomials $\bm P^{l}(\bm c)$ allows us to conveniently express the linearized collision operator in terms of $M_b^{ls}$ moments, the basis transformation of \cref{eq:basistransf} is needed to cast the velocity dependence of $\left< C(f_M \delta f_k)\right>$ in a Hermite-Laguerre polynomial basis, and to calculate its velocity moments.}

We now use the expression of $\left< C(f_M \delta f_k)\right>$ contained in \cref{eq:jiheldfmf1} and inject it in \cref{eq:gyromoments}.
By defining the fluid moments $A_{eb}^{lts}$ as
\begin{equation}
    A_{eb}^{lts} = \int v^l L_t^{l+1/2}(v^2) f_M \nu_{b}^{ls,0}(v)dv,
\end{equation}
and $B_{eb}^{lts}$ as
\begin{equation}
    B_{eb}^{lts} = \int v^l L_t^{l+1/2}(v^2) f_M \nu_{b}^{0,ls}(v)dv,
\end{equation}
the resulting collision operator moments $C^{pj}$ can be written as
\begin{align}
    C^{pj}=\sum_b\sum_{s=0}^{\infty}\sum_{l=0}^{p+2j}\sum_{t=0}^{j+\floor{p/2}}&\frac{(T^{-1})^{lt}_{pj} 2^l (l!)^2}{(2l)!\sigma_s^l\sqrt{2^p p!}}\frac{\nu_{b}}{(2l+1)}
    {\left(M_e^{ls} A_{eb}^{lts}+M_b^{ls} B_{eb}^{lts}\right)},
\label{eq:cabpj}
\end{align}
where we introduce the inverse transformation coefficients
\begin{equation}
    (T^{-1})_{pj}^{lt}=T_{lt}^{pj} \frac{\sqrt{\pi}2^p p!(l+1/2)t!}{(t+l+1/2)!}.
\end{equation}
The analytical expressions for $A_{eb}^{lts}$ and $B_{eb}^{lts}$ suitable for numerical implementation are given in \citet{Ji2006}.
The moments of the collision operator, $C^{pj}$, correspond to the ones derived in \citet{Jorge2018} for the study of drift-waves, and can also be obtained by linearizing the electron collisional moments presented in \citet{Jorge2017}.

Besides the Coulomb collision operator, the Hermite-Laguerre expansion described above can be advantageously applied to  describe other collision operators.
We consider here the Lenard-Bernstein \citep{Lenard1958}, the Dougherty \citep{Dougherty1964}, and the electron-ion collision operators that are used for comparison with the {full} Coulomb one.
The Lenard-Bernstein and Dougherty operators are implemented in a number of advanced kinetic codes \citep{Nakata2016,Grandgirard2016,Loureiro2015,Pan2018}, and are frequently used to introduce collisional effects in weakly collisional plasmas \citep{Zocco2011,Zocco2015,Shi2017,Mandell2018}.
Therefore, a comparison between the Coulomb and the Lenard-Bernstein and Dougherty operators, even in simplified systems such as the case of EPW, is important to determine the accuracy and validity of these operators.
The Lenard-Bernstein collision operator $C_{LB}(f)${, first derived in 1891 by L. Rayleigh \citep{Wax1954}} is of the Fokker-Planck type.
It conserves particle number and satisfies the H-theorem, and it can be written as \citep{Lenard1958}
\begin{equation}
    C_{LB}(f) = \nu \frac{\partial}{\partial \mathbf v} \cdot \left(\mathbf v f + \frac{v_{th}^2}{2}\frac{\partial f}{\partial \mathbf v}\right).
\label{eq:LBoperator}
\end{equation}
This operator can be derived from the Fokker-Planck equation, \cref{eq:caa}, by assuming $(m/m_b)\partial_{\mathbf v} H_b = \mathbf v$ and $\partial_{\mathbf v}\partial_{\mathbf v}G_b  = -\mathbf I v_{th}^2/2$ with $\mathbf I$ the identity matrix.
By projecting the Lenard-Bernstein operator onto a Hermite-Laguerre basis according to \cref{eq:projectioncoll}, one obtains
\begin{equation}
    C_{LB}^{pj} = - \nu (p+2j) N^{pj}.
\label{eq:clb}
\end{equation}
Equation (\ref{eq:clb}) can then be used in the moment-hierarchy equation \cref{eq:momenthierarchy}, yielding
\begin{align}
    i \frac{\partial}{\partial t} N^{pj} &= \sqrt{\frac{p+1}{2}}N^{p+1 j}+\sqrt{\frac{p}{2}}N^{p-1 j}+ \frac{N^{00}}{\alpha_D}\frac{\delta_{p,1}\delta_{j,0}}{\sqrt{2}}-i\nu (p+2j) N^{pj}.
\label{eq:lbmomenthierarchy}
\end{align}
The linearized Dougherty collision operator $C_D(f)$, on the other hand, adds the necessary field-particle collisional terms to the Lenard-Bernstein operator in order to provide momentum and energy conservation properties. Namely, it sets $(m/m_b)\partial_{\mathbf v} H_b = \mathbf v-\mathbf u$, with $\mathbf u = \int \mathbf v f dv_z dv_\perp^2 d\varphi/n_0$, and $\partial_{\mathbf v}\partial_{\mathbf v}G_b  = -\mathbf I T/m_a$ with $T=\int m (\bm v-\bm u)^2 f dv_z dv_\perp^2 d\varphi/(3 n_0)=(\sqrt{2}N^{20}-2N^{01})/3$.
{The Hermite-Laguerre moments of the linearized Dougherty collision operator $C_D^{pj}$ are given by
\begin{equation}
    C_{D}^{pj}=-\nu \left[(p+2j) N^{pj} - N^{10} \delta_{p1}\delta_{j0}+T(\sqrt{2}\delta_{p0}\delta_{j1}-2\delta_{p2}\delta_{j0})\right],
\end{equation}
}
{yielding} the moment-hierarchy equation
\begin{align}
    i \frac{\partial}{\partial t} N^{pj} &= \sqrt{\frac{p+1}{2}}N^{p+1 j}+\sqrt{\frac{p}{2}}N^{p-1 j}+ \frac{N^{00}}{\alpha_D}\frac{\delta_{p,1}\delta_{j,0}}{\sqrt{2}}\nonumber\\
    &-i\nu \left[(p+2j) N^{pj} - N^{10} \delta_{p1}\delta_{j0}+T(\sqrt{2}\delta_{p0}\delta_{j1}-2\delta_{p2}\delta_{j0})\right].
\label{eq:dmomenthierarchy}
\end{align}
{We note that the moment-hierarchies with the Lenard-Bernstein or the Dougherty collision operator, \cref{eq:lbmomenthierarchy} and \cref{eq:dmomenthierarchy}, respectively, do not couple different Laguerre moments and, therefore, one can focus on obtaining the coefficients $N^{p0}$ for solving the moment-hierarchy.}

The Hermite-Laguerre expansion procedure can also be applied to the electron-ion operator $C_{ei}(f)$ that is often used for EPW studies \citep{Epperlein1992,Banks2016}, that is
\begin{equation}
    C_{ei}(f) = \frac{\nu}{2 v^3}\frac{\partial}{\partial \xi}\left[\left(1-\xi^2\right)\frac{\partial f}{\partial \xi}\right],
\label{eq:cei}
\end{equation}
with $\xi = v_z/v$.
This operator describes the pitch-angle scattering of electrons due to collisions with ions.
By projecting \cref{eq:cei} into a Hermite-Laguerre basis, we obtain \citep{Jorge2017}
\begin{equation}
\begin{split}
    C_{ei}^{pj} = -\frac{\nu}{8 \pi^{3/2}}&
    \sum_{l=0}^{p+2j}\sum_{f=0}^{j+\floor{p/2}}
    \frac{{\l(T^{-1}\r)}_{pj}^{lf}}{\sqrt{2^p p!}}\sum_{s=0}^\infty A_{ei}^{lf,s} M_e^{ls},
\end{split}
\label{eq:ceipj}
\end{equation}
where the $A_{ei}^{lf,s}$ coefficients are given by
\be
    \begin{split}
        A_{ei}^{lf,s}=&\frac{l(l+1)}{l+1/2}\frac{2^l (l!)^2}{(2l)!} \sum_{m=0}^f \sum_{n=0}^s \frac{L_{fm}^l L_{sn}^l}{\sqrt{\sigma_s^l}} {(l+m+n-1)!}.
    \end{split}
\ee

As an aside, we note that previous studies on EPW have shown that the solutions of the linearized Boltzmann equation are, in fact, sensitive to the discretization method used.
{For example, it was shown that finite-difference methods, when applied to the problem of EPW, produce a number of numerical, non-physical modes with a rather small damping rate that do not lie in the vicinity of the collisionless solutions, even for weak collisionalities and a very high resolution \citep{Bratanov2013}}.
{On the other hand, {a} discretization scheme based on a Hermite-Laguerre polynomial decomposition yields a large number of roots that lie in the vicinity of the collisionless solution. Since} previous EPW studies using the electron-ion collision operator have been performed using a discretization of the distribution function into a set of Legendre polynomials, i.e., \citep{Epperlein1992,Brantov2012,Banks2016}
\begin{equation}
    \left<\delta f_k\right> = \sum_{l=0}^\infty a_l(v) P_l(\xi),
\label{eq:legendreexp}
\end{equation}
as a test of our approach, we compare {in \cref{sec:eigspec}} our results with the Legendre decomposition in \cref{eq:legendreexp}.
By projecting the Boltzmann equation, \cref{eq:boltzmanneq}, with an electron-ion collision operator into a Legendre basis, \cref{eq:legendreexp}, the following moment-hierarchy equation is obtained
\begin{equation}
    \frac{i}{v}\frac{d a_l(v)}{dt}  =  \frac{l}{2l-1} a_{l-1}(v)+\frac{l+1}{2l+3}a_{l+1}(v)+\frac{\delta_{l,1}}{\alpha_D}\int f_M v^2 a_0(v) dv -  \frac{i \nu}{v^4} l(l+1) a_l.
\label{eq:legmomenthi}
\end{equation}
A relation between the Hermite-Laguerre $N^{pj}$ and Legendre moments $a_l$ can be found by comparing \cref{eq:legendreexp,eq:gyrof}, yielding
\begin{equation}
    a_l(v)=\sum_{p=0}^\infty \sum_{j=0}^\infty \sum_{s=0}^{p+2j}\sum_{t=0}^{j+\floor{p/2}}(T^{-1})^{st}_{pj}\frac{N^{pj}}{\sqrt{2^p p!}}v^s L_t^{s+1/2}(v^2) \delta_{ls},
\end{equation}
and
\begin{equation}
    N^{pj} = \sum_{s=0}^{p+2j}\sum_{t=0}^{j+\floor{p/2}}\frac{(T^{-1})^{st}_{pj}}{\sqrt{2^p p!}(2l+1)}\int a_s(v) v^{s+2} L_t^{s+1/2}(v^2) dv.
\end{equation}

\section{Collisionless Dispersion Relation}
\label{sec:collisionless}

As a first step in the analysis of EPW, and for comparison with the results {in the presence of collisions}, we derive the EPW dispersion relation in the collisionless limit.
We first Fourier transform in time the collisionless limit of the moment-hierarchy equation, \cref{eq:momenthierarchy}, by imposing $\delta f_k \sim e^{(\gamma+i \omega) t}$, obtaining
\begin{align}
    i (\gamma+i \omega) N^{pj} &= \sqrt{\frac{p+1}{2}}N^{p+1 j}+\sqrt{\frac{p}{2}}N^{p-1 j}+ \frac{N^{00}}{\alpha_D}\frac{\delta_{p,1}\delta_{j,0}}{\sqrt{2}}.
\label{eq:collessmomenthierarchy}
\end{align}
A closed form solution of the collisionless moment-hierarchy in \cref{eq:collessmomenthierarchy} can be obtained by dividing the Boltzmann equation, \cref{eq:boltzmanneq}, by the resonant $i \gamma-v_z$ factor, multiplying it by the Hermite-Laguerre polynomial basis functions and, finally, integrating it over velocity space.
This yields
\begin{align}
    {N^{pj}}&=-\frac{N^{00}}{\alpha_D}\left[-i(\gamma+i\omega)\frac{(-1)^p}{\sqrt{2^p p!}}Z^{(p)}\left(\omega-i \gamma\right) + \delta_{p,0}\right]\delta_{j,0},
\label{eq:npjepw}
\end{align}
where $Z^{(p)}$ is the $p$th derivative of the plasma dispersion function $Z^{(0)}$, defined by
\begin{equation}
    Z^{(p)}(u)=\frac{(-1)^p}{\sqrt{\pi}}\int_{-\infty}^{\infty}\frac{H_p(x)e^{-x^2}}{x-u}dx.
    \label{eq:derznint}
\end{equation}
By setting $(p,j)=(0,0)$ in \cref{eq:npjepw},  the collisionless dispersion relation is found
\begin{equation}
    D=1+\alpha_D-i(\gamma+i \omega) Z(\omega-i\gamma)=0.
\label{eq:collisionless}
\end{equation}
{Alternatively, \cref{eq:collisionless} can be derived from the collisionless limit of the Boltzmann equation, \cref{eq:boltzmanneq}, upon division by the factor $i\gamma-\omega-v_z$ and integration with respect to $v_z$.}
The numerical solution of \cref{eq:collisionless} is obtained by discretizing $\gamma$ and $\omega$ into a two-dimensional $[\omega,\gamma]$ grid, evaluating $D$ on the grid, and storing the values where $[Re(D),Im(D)]$ vanishes.
To evaluate $Z$, we make use of the identity $Z(x)=i \sqrt{\pi} e^{-x^2}\text{erfc}{(-i x)}$ with $\text{erfc}{( x)}=1-\text{erf}(x)$ and $\text{erf}(x)$ the error function, and use the algorithm developed in \citet{Gautschi1970} to numerically compute $\text{erf}(x)$ for complex arguments.

\begin{figure}
    \centering
    \includegraphics[width=.80\textwidth]{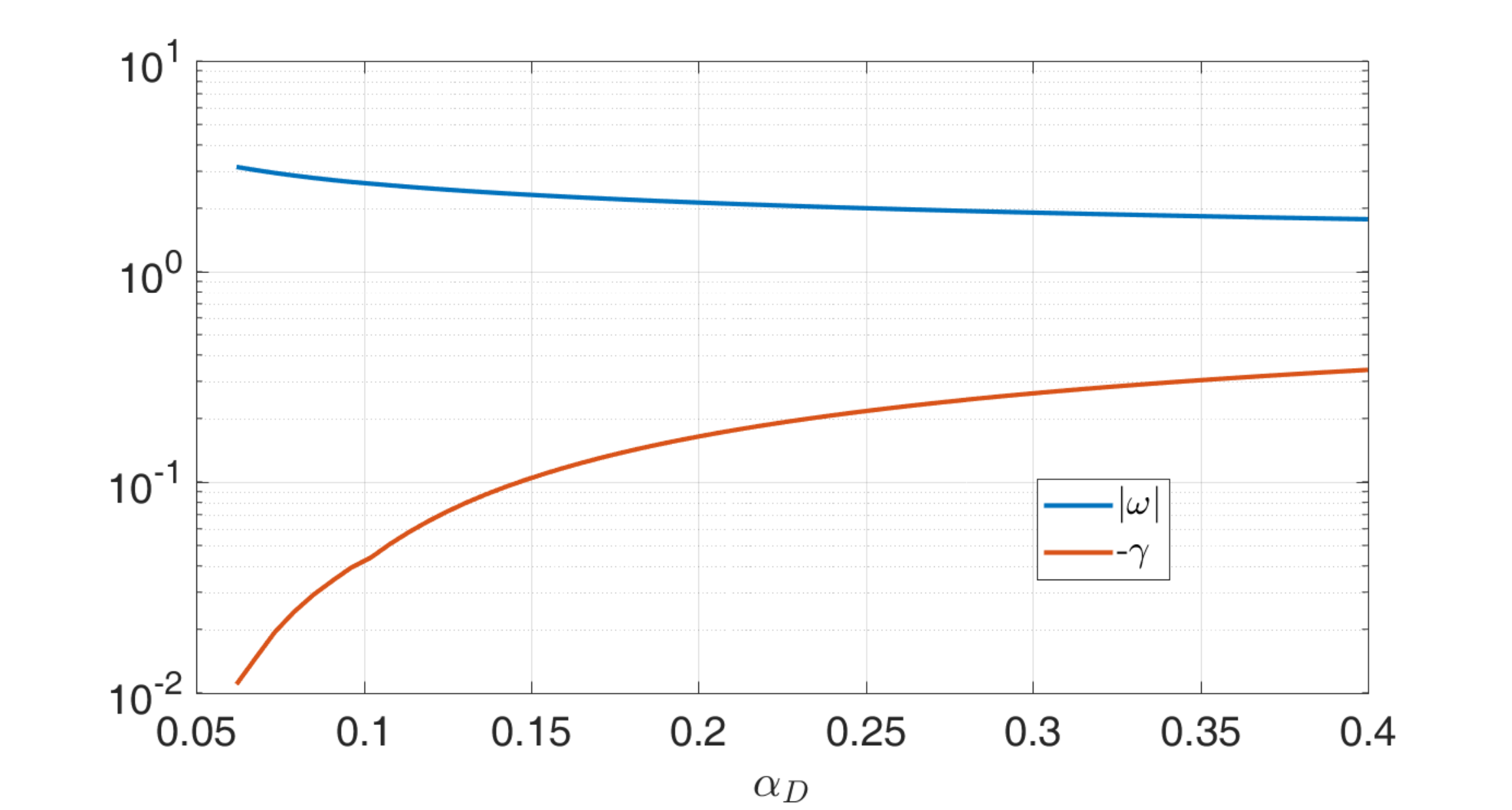}
    \caption{Collisionless frequency {(blue line)} and damping rate {(red line)} of the least damped {solution of the collisionless dispersion relation, \cref{eq:collisionless},} as a function of $\alpha_D$, for $0.05<\alpha_D<0.4$.}
    \label{fig:colless_alphad}
\end{figure}

The $\alpha_D$ dependence of the least damped solution of \cref{eq:collisionless} is shown in \cref{fig:colless_alphad}, where both its damping rate $\gamma$ and frequency $\omega$ are seen to be monotonic functions of $\alpha_D$, which is in agreement with previous EPW studies \citep{Banks2017}.
In the following, without loss of generality and similarly to previous studies of collisional damping of EPW \citep{Banks2016,Banks2017}, we select the value of $\alpha_D=0.09$ when fixed $\alpha_D$ studies are performed, which corresponds to $k \lambda_D = \sqrt{\alpha_D}=0.3$.
{While} this {value of} $\alpha_D$ is typical for EPW driven by stimulated Raman scattering \citep{Brunner2004,Winjum2013}{, we add that the particular choice of $\alpha_D$ has no quantitative impact on the conclusions we draw.}
\section{Temporal Evolution of EPW}
\label{sec:simresults}

\begin{figure}
    \centering
    \includegraphics[width=0.8\textwidth]{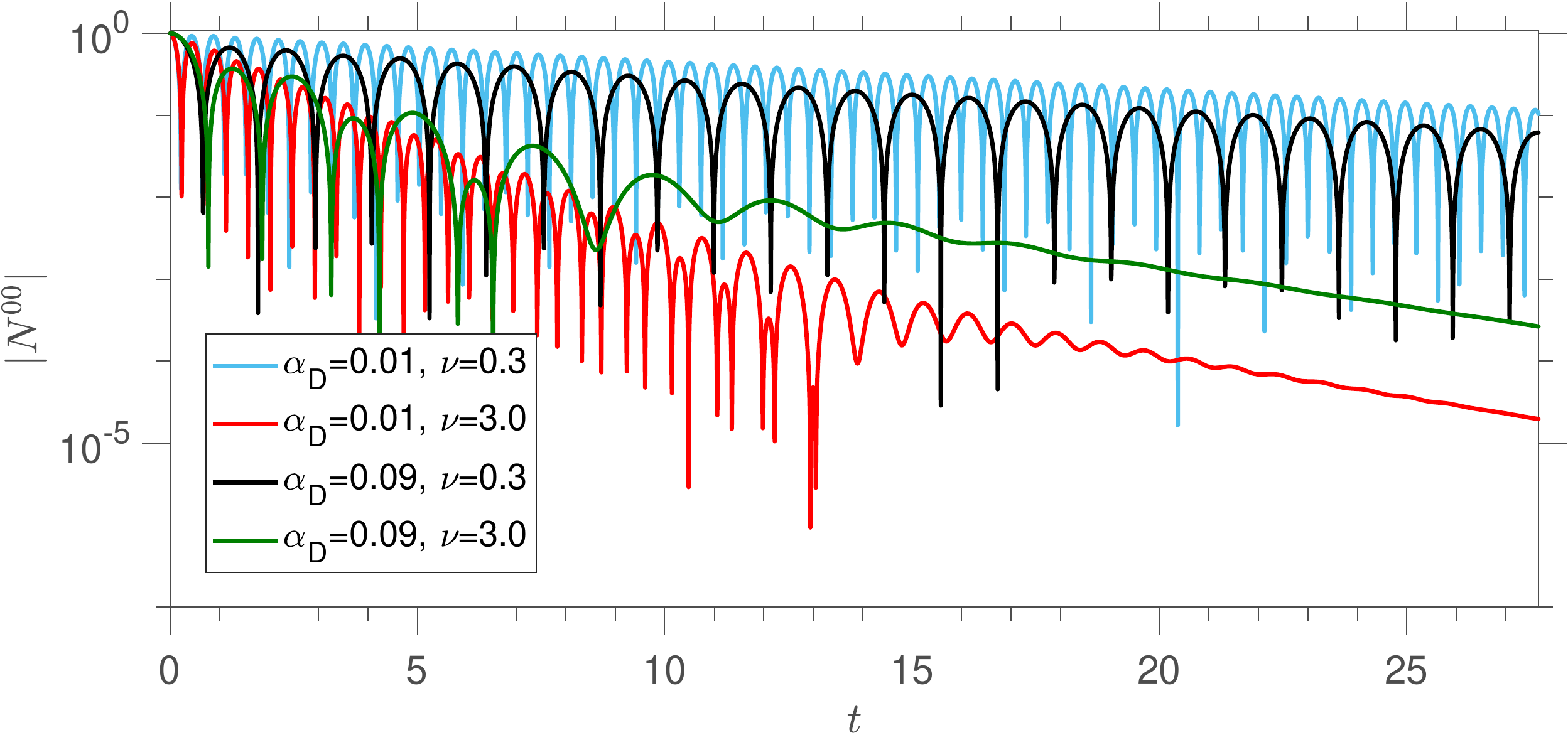}
    \caption{Time evolution of the absolute value of $N^{00}$ using a truncation with $(P,J)=(18,2)${, evaluated using the full linearized Coulomb collision operator}. Different values of $\nu$ and $\alpha_D$ are shown.
    }
    \label{fig:timetraces}
\end{figure}

In this section, the moment-hierarchy equation, \cref{eq:momenthierarchy}, is solved numerically as a time-evolution problem {using an implicit variable-step variable-order solver that employs backward differentiation formulas \citep{Shampine2002} with a maximum time-step of $10^{-6} k v_{th}$}.
For the numerical solution, the moment-hierarchy is truncated at a maximum Hermite-Laguerre index $(P,J)$ by setting
\begin{equation}
    N^{pj}=0,~\text{for}~(p,j)>(P,J).
\label{eq:trunc}
\end{equation}
We consider as initial condition $N^{pj}(t=0)=\delta_{p0}\delta_{j0}$, such that the perturbed density and electrostatic potential are initially excited, while higher moments of the distribution function are set to zero.
{The initial perturbation of the velocity distribution function is then taken to be a Maxwellian in velocity space.}
The temporal evolution of $N^{00}$ (and therefore of $\phi$) is shown in \cref{fig:timetraces} {for different collisionalities and $\alpha_D$ values}.
In this section, we focus on {the} oscillating initial phase {of \cref{fig:timetraces}}, where the EPW dominate the dynamics.
We fit the amplitude of $N^{00}$ to an exponentially damped sinusoidal wave with real frequency $\omega$ and damping rate $\gamma${, taking into account a minimum of three oscillation periods}.
The later phase, where a purely damped behavior {is observed} at higher collisionalities due to the presence of an entropy mode, is investigated in \cref{sec:timeevol}.

\begin{figure}
    \centering
    \includegraphics[width=0.86\textwidth]{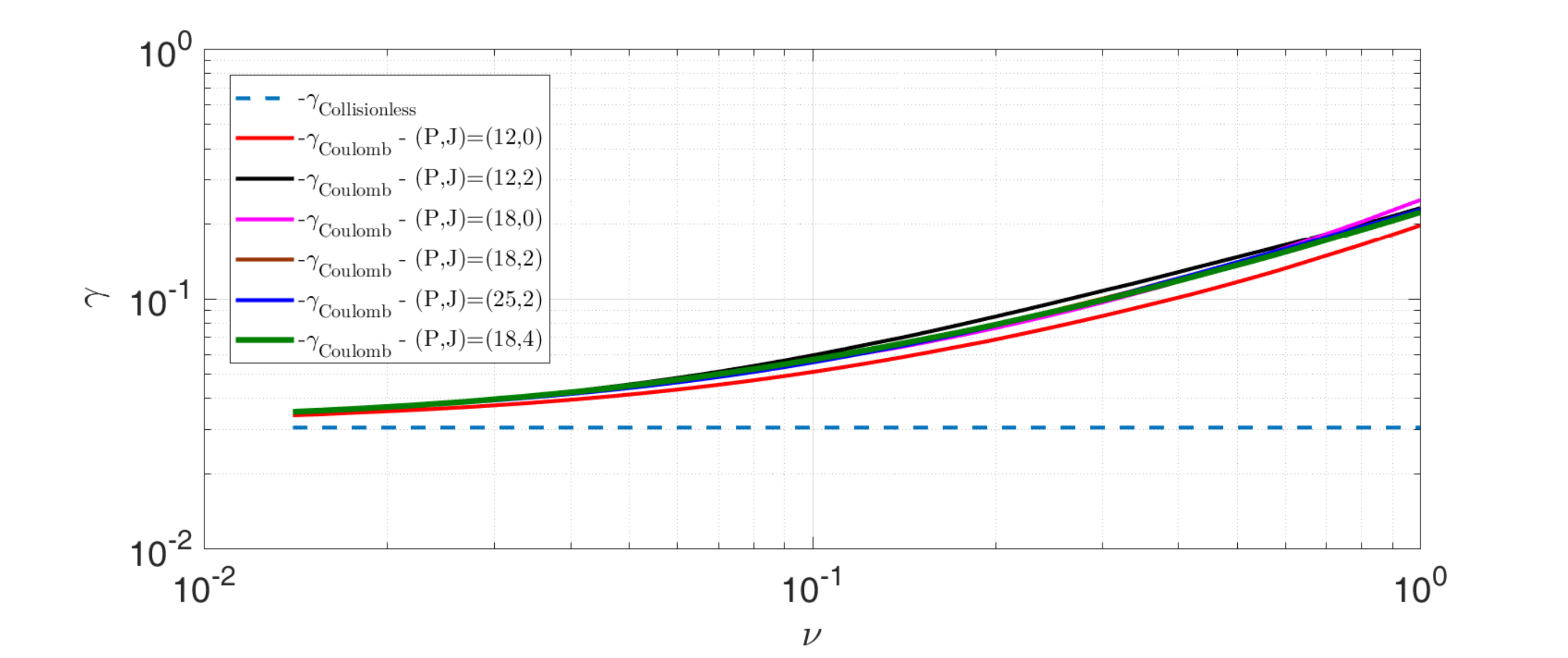}
    \caption{Comparison of the {normalized} damping rate $\gamma$ for $\alpha_D=0.09$ {as a function of the normalized collision frequency} considering a truncation at different values of $(P,J)$ (solid lines) {and using the full linearized Coulomb collision operator}.
    The collisionless least damped Landau solution is shown for comparison (dashed blue line).
    {All frequencies are normalized to $k v_{th}$.}
    }
    \label{fig:convstudy}
\end{figure}

\begin{figure}
    \centering
    \includegraphics[width=0.49\textwidth]{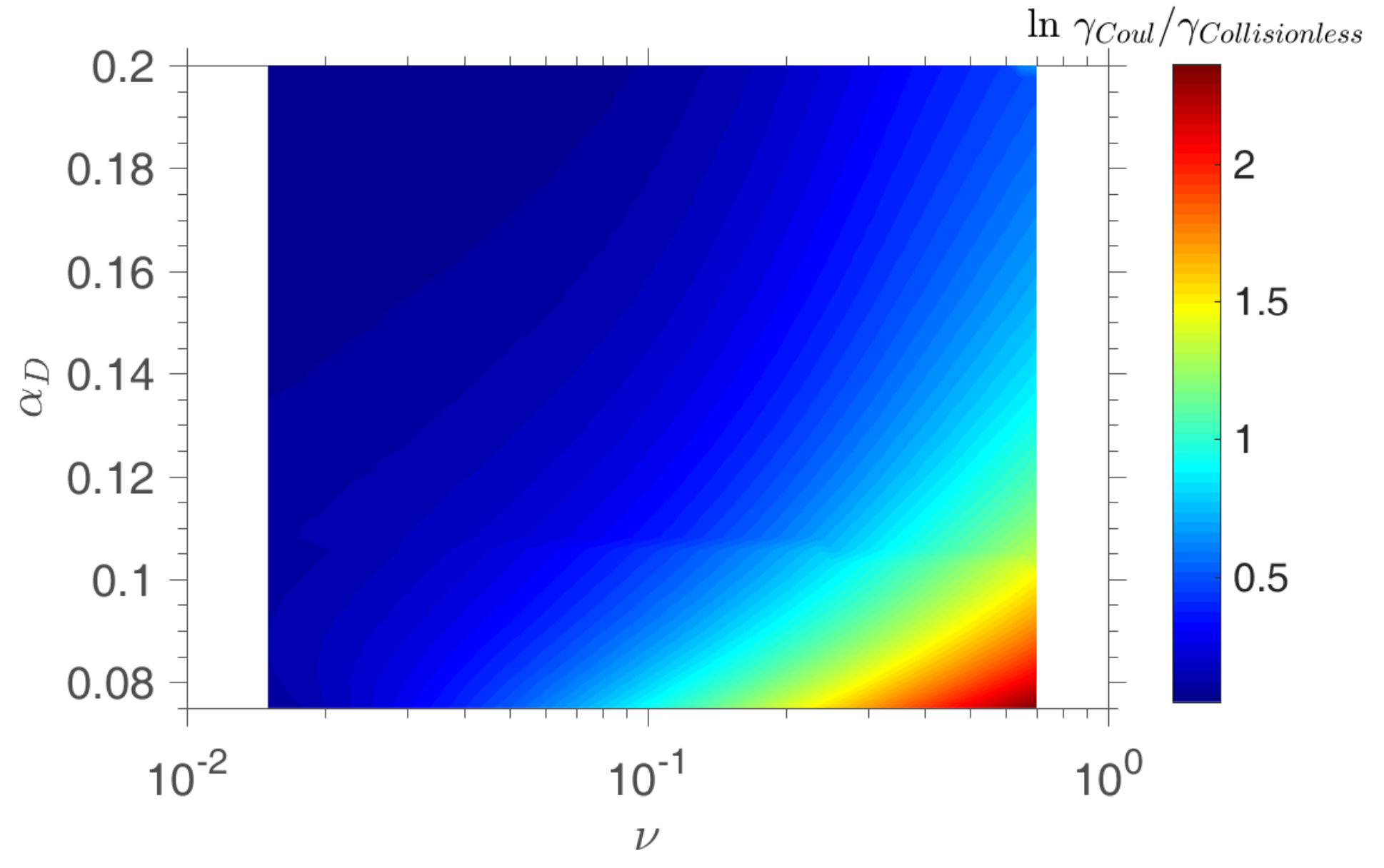}
    \includegraphics[width=0.49\textwidth]{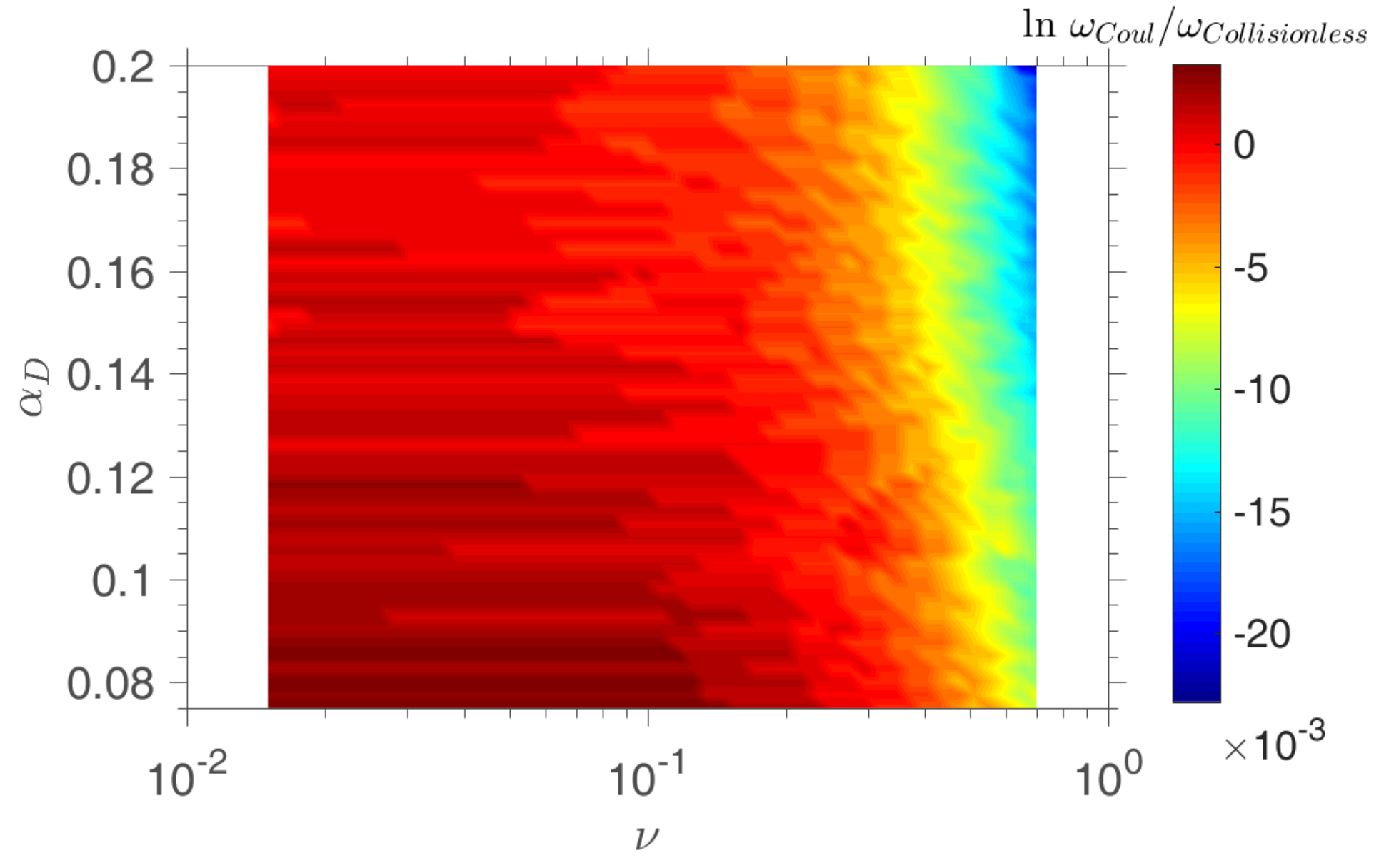}
    \caption{Damping rate $\gamma$ (left) and oscillation frequency $\omega$ (right) of the electron-plasma wave obtained from the moment-hierarchy equation, \cref{eq:momenthierarchy}, as a function of $\alpha_D$ and $\nu$ for $(P,J)=(18,2)$.
    {The full linearized Coulomb collision operator is considered.}}
    \label{fig:momhiersol}
\end{figure}

A convergence study with the truncation indices $(P,J)$ is shown in \cref{fig:convstudy}.
{Convergence is observed for $(P,J) = (18,2)$ in the range of collisionalities and $\alpha_D$ investigated  {(a variation of less than 3\% is observed between damping rates evaluated with a truncation at $(P,J) = (18,2)$ and a truncation at higher values of $P$ and $J$)}.}
The values of {oscillation frequency} $\omega(\alpha_D,\nu)$ and {damping rate} $\gamma(\alpha_D,\nu)$ obtained as a fit of the initial damping phase {are} shown in \cref{fig:momhiersol} for {$0.075<\alpha_D<0.2$ and $0.015<\nu<0.7$} using the Coulomb collision operator, where a truncation at $(P,J)=(18,2)$ is used.
{Such values of $(P,J)$ are in line with the estimate found in \citet{Jorge2018} which, by underestimating the effect of the collisional term $C^{pj}$ in the moment-hierarchy equation leads to $P\sim 4/\sqrt 2 \nu$ and $J\sim 2$.
For $\nu \sim 0.1$, this yields $(P,J)\sim(30,2)$.}
The largest deviation of the damping rate from the collisionless {case} is seen to occur for large values of collisionality and small $\alpha_D$. This is expected, as for large $\nu$ and small $\alpha_D$ the collisional fluid limit is retrieved.
{The dependence on $\alpha_D$} may be attributed to the decreasing magnitude of the Landau damping rate for decreasing $\alpha_D$ (see \cref{fig:colless_alphad}).
This makes the ratio between the collisional and the collisionless damping rates increasingly larger.
{Finally, we remark that the presence of several competing eigenmodes in the {initial transients of the} temporal evolution of $N^{00}$ contribute to the presence of a transition at $\alpha_D \sim 0.1$ visible in \cref{fig:momhiersol}.}

\begin{figure}
    \centering
    \includegraphics[width=0.7\textwidth]{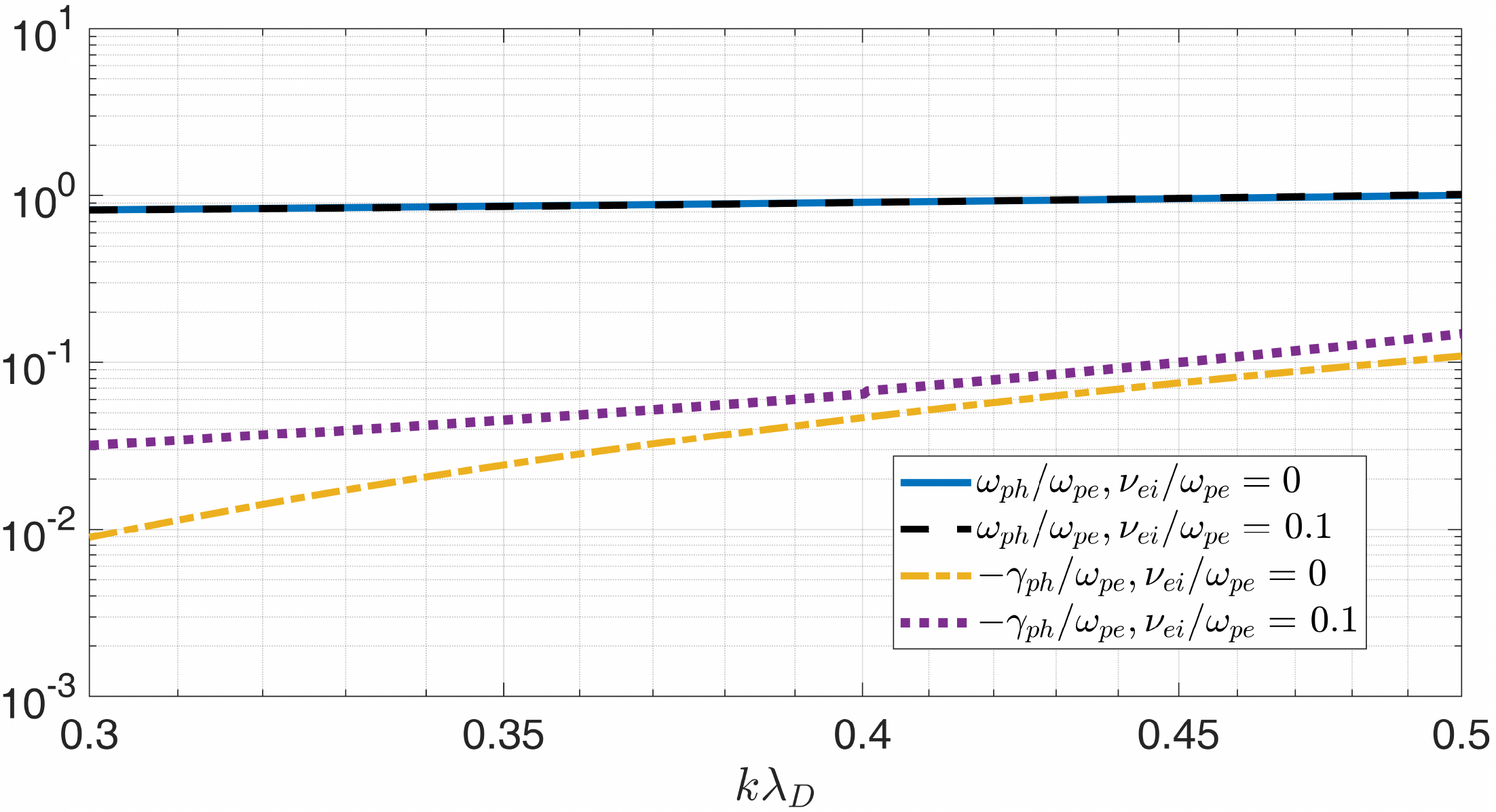}
    \caption{{Oscillation frequency $\omega_{ph}=\omega k v_{th}$ and damping rate $\gamma_{ph}=\gamma k v_{th}$ of EPW in physical units, normalized to the plasma frequency $\omega_{pe}$ for the collisionless $\nu_{ei}/\omega_{pe}=0$ and collisional $\nu_{ei}/\omega_{pe}=0.1$ case.}}
    \label{fig:jackson}
\end{figure}

\begin{figure}
    \centering
    \includegraphics[width=0.75\textwidth]{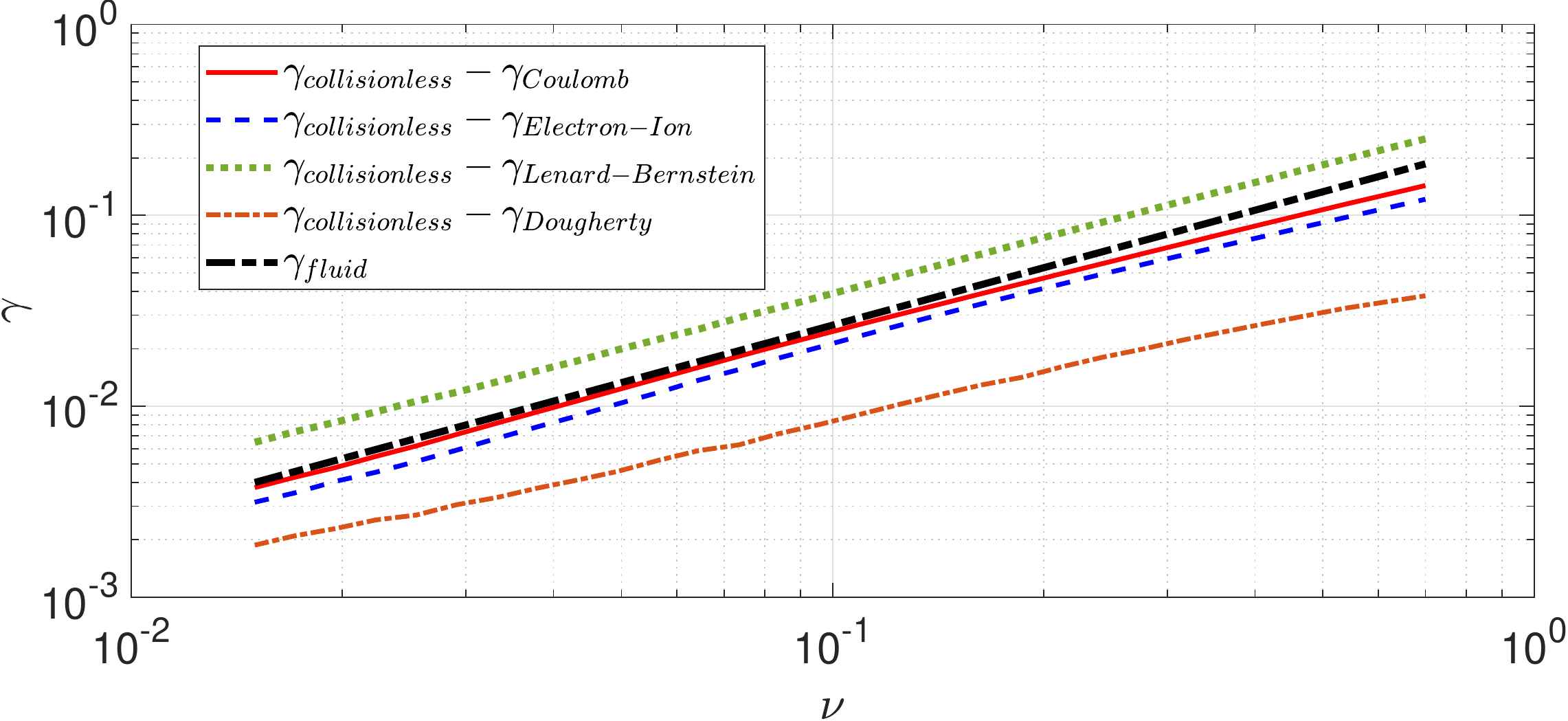}
    \caption{Difference between the collisionless damping rate $\gamma$ with the one resulting from the moment-hierarchy equation, \cref{eq:momenthierarchy}, with the full Coulomb, Lenard-Bernstein, Dougherty, and electron-ion collision operators at $\alpha_D=0.09$ and $(P,J)=(18,2)$. {The collisional damping rate $\gamma_{fluid}=-0.266 \nu$ obtained from a fluid description is also shown for comparison.}}
    \label{fig:coulLBcomp}
\end{figure}

{Following a long standing tradition established by Jackson \citep{Jackson1960}, and used in consequent EPW studies \citep{Opher2002}, we also display the frequency and the damping rate of EPW normalized to the plasma frequency $\omega_{pe}=v_{th}/\lambda_D$ in \cref{fig:jackson} as a function of $k \lambda_D$ for fixed $\nu_{ei}/\omega_{pe}$.
We show the results for the collisionless $\nu_{ei}/\omega_{pe}=0$ and the collisional $\nu_{ei}/\omega_{pe}=0.1$ case.
Consistently with \cref{fig:momhiersol}, it is observed that $\omega/\omega_{pe}$ has a negligible relative variation with the collision frequency when compared with the relative variation of the damping rate.
Furthermore, as observed in \citet{Opher2002}, the damping rate of EPW with finite collision frequency is seen to approach the collisionless result for high values of $k \lambda_D$.
}
{
A further comparison of the collisional component of the damping rate $\gamma_{coll}$, obtained with different collision models, is shown in \cref{fig:coulLBcomp},  where $\gamma_{coll} = \gamma - \gamma_{collisionless}$ with $\gamma$ the total damping rate and $\gamma_{collisionless}$ the collisionless Landau damping.
Results considering Lenard-Bernstein, Dougherty, electron-ion, and the full Coulomb collision operators are shown.
}
{We note that when the Lenard-Bernstein and the Dougherty operator are considered, only self-collisions are taken into account, and with the electron-ion operator only unlike-particle collisions are included.}
In general, the Coulomb operator yields a damping rate smaller than the Lenard-Bernstein and larger than the Dougherty one, with deviations of up to $50\%$ between different operators.
The use of an electron-ion collision operator is preferable since it yields  damping rates and frequencies similar, just slightly lower, than the Coulomb operator.
{The collisional damping rate, $\gamma_{fluid}=-0.532\nu/2$, obtained from a fluid description using the Braginskii equations \citep{Banks2017}, is also shown for comparison.}
{We remark that the results in \cref{fig:coulLBcomp} for the collisional component of the damping rate of the fluid, purely e-i collisions, and the full Coulomb operator are in close agreement with the findings of \citet{Banks2017}.}

For $\nu\ll 1$, it is seen that the damping rates of all solutions approach the collisionless limit regardless of the collision operator used.
When $\nu$ is increased, \cref{fig:coulLBcomp} shows that the differences between the collision operators {still persist}.
This allows us to draw arguments for the difference between the different collision operators by using a low number of moments.
The lowest order {particle conservation} $C^{00}=0$ and {collisional friction} $C^{10}=-\nu N^{10}$ moments $C^{pj}$ are the same between the Coulomb, Lenard-Bernstein, and electron-ion collision operators, while the Dougherty operator has $C_D^{10}=0$.
This effectively reduces the damping rate {evaluated with the Dougherty operator with respect to the Coulomb case}, as seen in \cref{fig:coulLBcomp}.

On the other hand, only the Coulomb, the Dougherty, and the electron-ion collision operators are energy conserving, i.e., satisfying $(1/2)\int m (v_z^2+v_\perp^2) C(f) d \mathbf v=0$ or, equivalently, $C^{20}=\sqrt{2}C^{01}$, while the Lenard-Bernstein operator does not conserve energy.
In fact, the energy moments are given by $C_{LB}^{01}=-\nu 2 N^{01}$ and $C_{LB}^{20}=-\nu 2 N^{20}$, yielding $C_{LB}^{01}=C_{LB}^{20}$.
However, despite the additional conservation properties, the agreement of the Dougherty operator is rather poor, as seen in \cref{fig:coulLBcomp}.
We conclude therefore that the presence of additional momentum and energy conserving terms in the Dougherty operator with respect to the Lenard-Bernstein operator does not yield a damping rate closer to the Coulomb one.
This was also pointed out in \citet{Jorge2018}, where a similar framework was used to derive the growth rate of the drift-wave instability.
%

\section{Entropy Mode}
\label{sec:timeevol}

\begin{figure}
    \centering
    \includegraphics[width=0.49\textwidth]{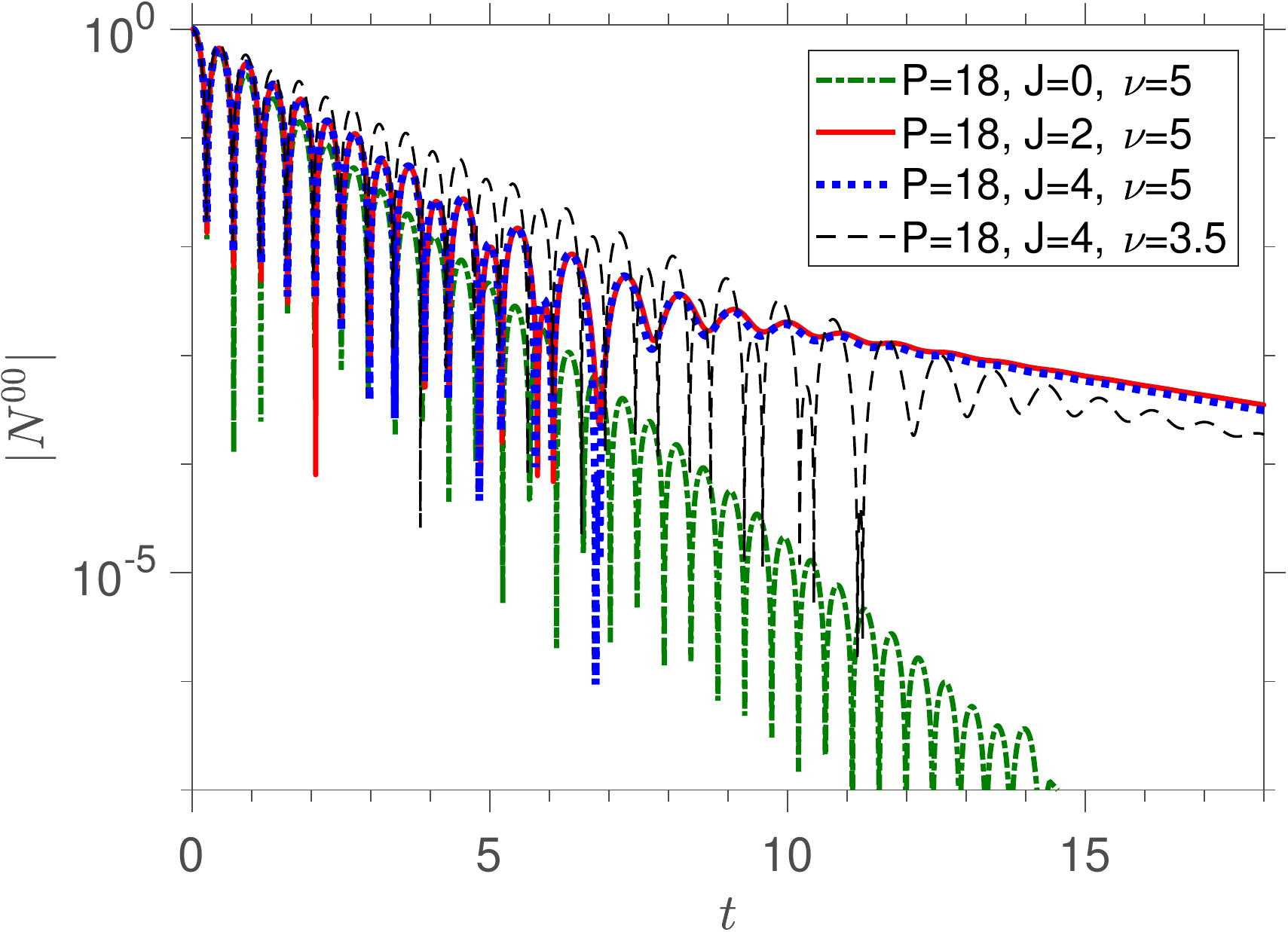}
    \includegraphics[width=0.49\textwidth]{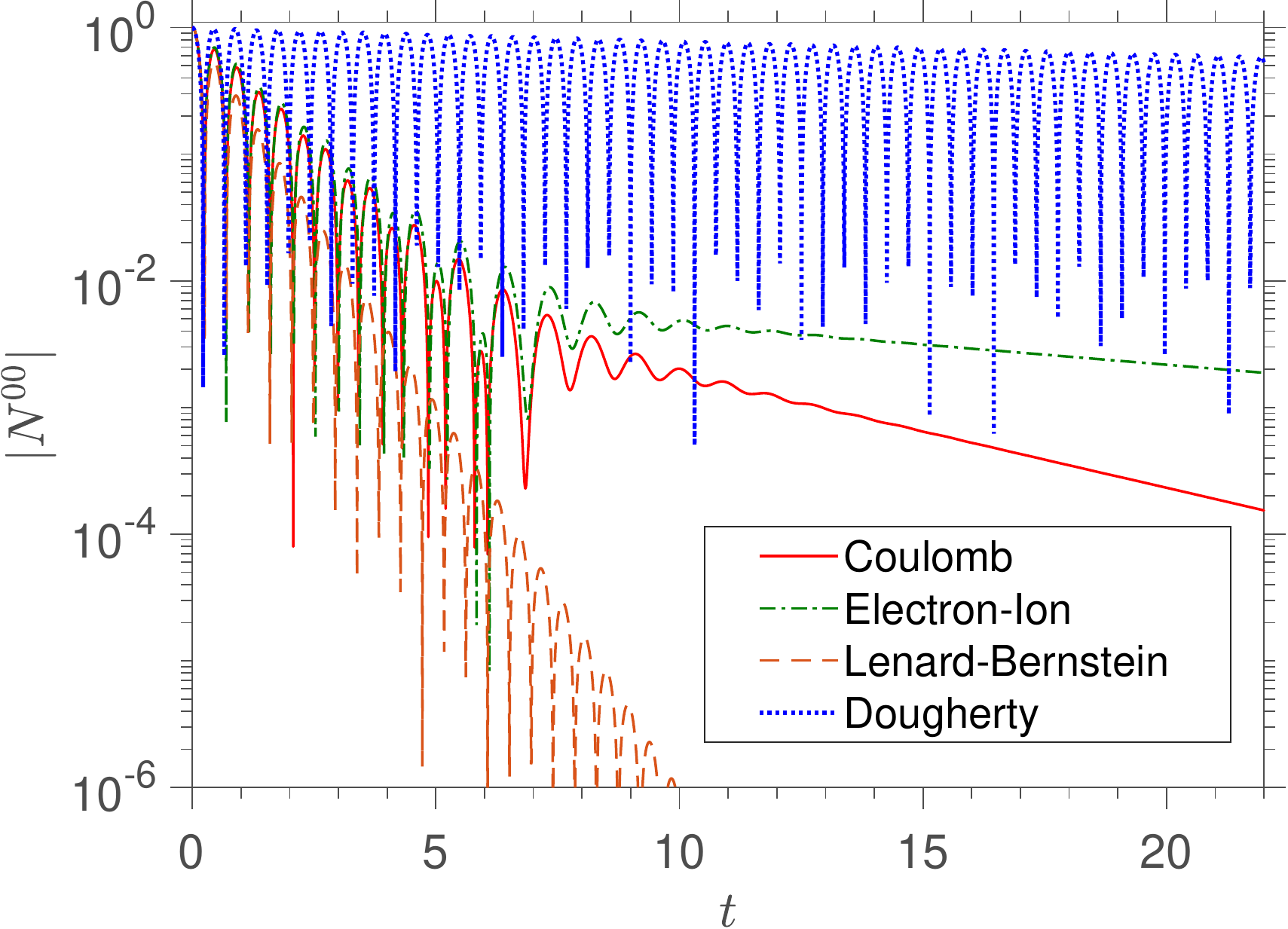}
    \caption{Time evolution of the absolute value of $N^{00}$ with $\nu=5$ and $\alpha_D=0.01$. Left: convergence study with a full-Coulomb operator and $(P,J)=(18,0)$ (green), $(P,J)=(18,2)$ (red), and $(P,J)=(18,4)$ (blue). {The time evolution of $N^{00}$ with $\nu=3.5$ and $(P,J)=(18,4)$ is also shown for comparison (black)}. Right: truncation at $(P,J)=(18,2)$ using the full-Coulomb collision operator (red), electron-ion collisions only (green), the Lenard-Bernstein (orange) and the Dougherty (blue) collision operators.}
    \label{fig:timetracecoul2}
\end{figure}

We focus on the latter stage of the time evolution of $N^{00}$ shown in \cref{fig:timetraces,fig:timetracecoul2}, where a purely damped behavior is found at high collisionalities.
In order to enhance the role of the zero-frequency mode, we consider the collision frequency $\nu=5$, while decreasing the role of Landau damping by setting $\alpha_D=0.01$ (the purely damped mode is not affected by the value of $\alpha_D$, if $\alpha_D\ll 1$).
{Indeed, the transition to a purely damped behaviour is seen to occur at times that decrease with the collision frequency [see \cref{fig:timetracecoul2} (black)].}
The resulting time traces of $|N^{00}|$ using a full-Coulomb collision operator are shown in \cref{fig:timetracecoul2} (left) for $(P,J)=(18,0)$, $(P,J)=(18,2)$, and $(P,J)=(18,4)$, while  time traces using the full-Coulomb, electron-ion, Lenard-Bernstein and the Dougherty collision operators with $(P,J)=(18,2)$ are shown in \cref{fig:timetracecoul2} (right).
We observe that for the Coulomb and electron-ion case, there is a transition to a purely damped mode at $t \simeq 7$ only when perpendicular velocity dynamics is introduced, $J\ge 2$.
{At the same time,} while for the Coulomb operator, the purely damped mode {that sets the late time evolution of the system in \cref{fig:timetracecoul2}} has a damping rate $\gamma \simeq -0.202$, the electron-ion collision operator yields a damping rate one order of magnitude smaller, $\gamma \simeq -0.024$.
This purely damped decay is not present when the Lenard-Bernstein or the Dougherty operators are considered.
We therefore conclude that in order to obtain the correct long-term behaviour of the Boltzmann equation, Coulomb self-collisions must be included in the description.

The long term behaviour observed in \cref{fig:timetracecoul2} is due to the presence of the entropy mode \citep{Banks2016}.
An analytical framework to model the entropy mode can be derived by noting that this is a purely damped mode with a damping rate much smaller than both the plasma  and the collision frequencies.
Previous studies on the collisional damping of EPW show that such a mode results from the effect of pitch-angle scattering at high-collisionality \citep{Banks2016}.
Considering that in the high-collisionality limit only the lowest order terms in the expansion of $\left<\delta f_k\right>$ in \cref{eq:gyrof} play a role and that, according to \cref{fig:timetracecoul2}, a finite perpendicular velocity-space resolution is an essential element for the entropy mode, we consider in the moment-hierarchy equation, \cref{eq:momenthierarchy}, the six lowest order Hermite-Laguerre expansion coefficients, namely $N^{00}, N^{10}, N^{20}, N^{30}, N^{01}$ and $N^{11}$ {with the ordering $N^{30}\sim N^{11} \sim \epsilon N^{20} \sim \epsilon N^{01}\sim\epsilon \phi$, with $\epsilon$ the small expansion parameter}
\begin{equation}
	\epsilon \sim \frac{1}{\nu} \sim \gamma,
	\label{eq:entordering1}
\end{equation}
{so that the particle mean-free path $\lambda_{mfp}=v_{th}/\nu_{ei}$ is small compared with typical wavelengths of the perturbed quantities, i.e., $k\lambda_{mfp}\ll 1$. Higher order moments are considered to be O$(\epsilon^2 \phi)$.}
{Charge neutrality is kept up to second order, i.e.,}
\begin{equation}
    \alpha_D \sim \epsilon^2.
\label{eq:hcdr}
\end{equation}
Using Poisson's equation, we find that density perturbations $N^{00}$ are negligible when compared with electrostatic fluctuations, namely
\begin{equation}
    \frac{N^{00}}{\phi} = -\alpha_D \ll 1.
\label{eq:noophiem}
\end{equation}
The moment-hierarchy equation, \cref{eq:momenthierarchy}, at $(p,j)=(0,0)$, shows that $N^{10}/N^{00} \sim \gamma$.
Together with the estimate in \cref{eq:noophiem}, this yields
\begin{equation}
    \frac{N^{10}}{\phi} \sim \gamma \alpha_D \ll 1
\label{eq:n10phiem}
\end{equation}
{Using the moment-hierarchy equation \cref{eq:momenthierarchy} and neglecting second order terms in the parallel $(p,j)=(2,0)$ and perpendicular $(p,j)=(0,1)$ temperature equations, we find that}
\begin{equation}
	{i \gamma N^{20} \simeq \sqrt{\frac{3}{{2}}}N^{30}-i \nu(0.45N^{01}+0.64N^{20}),}
	\label{eq:truncentropy1}
\end{equation}
{and}
\begin{equation}
	{i \gamma N^{01} \simeq \frac{N^{11}}{\sqrt{2}}-i \nu(0.32N^{01}+0.45N^{20}),}
	\label{eq:truncentropy2}
\end{equation}
{respectively.}
{The same procedure in the $(p,j)=(3,0)$ and $(p,j)=(1,1)$ moment equations yields}
\begin{equation}
	{0 \simeq \sqrt{\frac{3}{{2}}}N^{20}-i \nu(0.15N^{11}+1.03N^{30}),}
	\label{eq:truncentropy3}
\end{equation}
{and}
\begin{equation}
	{0 \simeq \frac{N^{01}}{\sqrt{2}}-i \nu(1.09N^{11}+0.15N^{30}),}
	\label{eq:truncentropy4}
\end{equation}
{respectively.}

As a consequence, the truncated moment-hierarchy equations, Eqs. (\ref{eq:truncentropy1})-(\ref{eq:truncentropy4}), yield the following dispersion relation
\begin{equation}
    \gamma^2 + 1.96 \gamma \left(\frac{1}{\nu}+0.49 \nu\right)+\frac{0.69}{\nu^2}+1.4 \times 10^{-5} \nu^2+0.88\simeq0
\label{eq:disprelentropymode}
\end{equation}
{that up to second order in $\epsilon$ yields the solutions $\gamma \simeq -0.96 \nu-1.04/\nu$ and $\gamma \simeq -0.92/\nu$.}
{The least damped solution,which is the one consistent with the ordering $\gamma \sim 1/\nu$  in \cref{eq:entordering1}, when applied to the $\nu=5$ case of \cref{fig:timetracecoul2}, leads to $\gamma \simeq -0.18$, which has a relative difference of $11\%$ with respect to the $\gamma = -0.202$ value obtained numerically.
}

We note that, with the same ordering above, a purely damped solution can also be obtained from the one-dimensional linearized Braginskii equations \citep{Braginskii1965}.
In this limit, in fact, the following linearized electron temperature equation is found
\begin{align}
    n_0 \frac{3}{2}\frac{\partial T_e}{\partial t}+\nabla_z \left(- \chi_{\parallel}^e \nabla_z T_e \right) \simeq 0
\label{eq:brageqt}
\end{align}
where $\chi_{\parallel}^e = 3.2 n_0 T_e/(m_e \nu k v_{th})$, with the Joule heating term proportional to $m_e/m_i$ neglected.
Equation (\ref{eq:brageqt}) yields the electron Braginskii entropy mode $\gamma \simeq - 1.1/\nu$, a value that is close to the estimate {above} based on the truncated moment-hierarchy equation.

{Finally, we remark that a purely damped mode is only observed in the temporal evolution of $N^{00}$ for values of $\nu \gtrsim 1$, while for $\nu \lesssim 1$ a transition from damped oscillations to a purely damped behaviour is not seen to occur for the range of values of $\alpha_D$ considered here even at later times.
The value of $\nu$ where a transition from collisional Landau damping to a purely damped entropy mode occurs after an initial transient is visible in the time evolution of $N^{00}$ can be estimated by balancing the damping rate of the collisional damping of EPW with the damping rate of entropy modes.
Estimating the former as $\gamma \simeq -0.03-0.26 \nu$ from \cref{fig:coulLBcomp}, and the latter as $\gamma \simeq -0.92/\nu$, the collision frequency at which the transition occurs is therefore estimated to be $\nu \simeq 1.8$, in agreement with the numerical results.
}

\section{Eigenvalue Spectrum}
\label{sec:eigspec}

We now compute the eigenmode spectra of EPW, and highlight the differences between the spectra of the full Coulomb, electron-ion, and Lenard-Bernstein operators using a Hermite-Laguerre decomposition, and the electron-ion operator using a Legendre polynomial decomposition.
{We note that subdominant and stable modes can be nonlinearly excited to finite amplitude \citep{Terry2006,Hatch2011,Pueschel2016,Hatch2016} and have a major role in nonlinear energy dissipation and turbulence saturation, affecting structure formation, as well as heat and particle transport.}
We note that both the Lenard-Bernstein and the Dougherty collision operators are seen to yield similar eigenmode spectra.
Therefore, we do not consider the Dougherty operator for this analysis.
To compute the Hermite-Laguerre EPW eigenmode spectrum, the moment-hierarchy equation, \cref{eq:momenthierarchy}, is truncated at a maximum index $(P,J)$, Fourier transformed in time, and the resulting eigenvalue problem solved numerically{, yielding the spectrum of solutions at arbitrary collisionality}.
In matrix form, this yields
\begin{equation}
    \mathbf A N = (\omega+i \gamma) N,
\end{equation}
where $N = [N^{00} N^{01} ...~N^{0J} N^{10} N^{11} ... N^{pj}~]$ is the moment vector and $\mathbf{A}$ the $(P+1)(J+1)\times (P+1)(J+1)$ matrix of moment-hierarchy coefficients of elements $A^m_n$ with $m$ and $n$ the row and column, respectively
{
\begin{align}
    A^{pJ+j}_{p'J+j'}=\sqrt{\frac{p+1}{2}}\delta_{p+1,p'}\delta_{j,j'}+\sqrt{\frac{p}{2}}\delta_{p-1,p'}\delta_{j,j'}+\frac{\delta_{p,1}\delta_{j,0}}{\sqrt{2}}\frac{\delta_{p',0}\delta_{j',0}}{\alpha_D}+iC^{pJ+j}_{p'J+j'},
\label{eq:amatrixeq}
\end{align}
}
{which can be written in matrix form as}
\begin{equation}
    A = \begin{bmatrix} 
    0                         & 0                      & \dots & 1/\sqrt{2} & 0                       & \dots \\
    0                         &i C_{01}^{01}            & \dots &i C_{10}^{01} & 1/\sqrt{2}+iC_{11}^{01} & \dots \\
    \vdots                    & \vdots                 & \dots & \vdots      & \vdots      & \dots \\
    (1+1/\alpha_D)/\sqrt{2} & iC_{01}^{10}            & \dots & iC_{10}^{10} & iC_{11}^{10} & \dots \\
    0                         & 1\sqrt{2}+C_{01}^{11} & \dots &i C_{10}^{11} & iC_{11}^{11} & \dots \\
    \vdots                    & \vdots                 & \dots & \vdots      & \vdots      & \dots 
    \end{bmatrix}.
\label{eq:amatrix}
\end{equation}
In \cref{eq:amatrixeq,eq:amatrix}, we have defined the collisional coefficients $C^{pj}_{st}$ in terms of the collisional moments $C^{pj}$ as $C^{pj}=\sum_{p',j'}C^{pj}_{p'j'}N^{p'j'}$ and $C^{pJ+j}_{p'J+j'}=C^{pj}_{p'j'}$.
The spectrum of $\gamma$ and $\omega$ is then found by computing the eigenvalues of the matrix $\mathbf A$.

\begin{figure}
    \centering
    \includegraphics[width=0.99\textwidth]{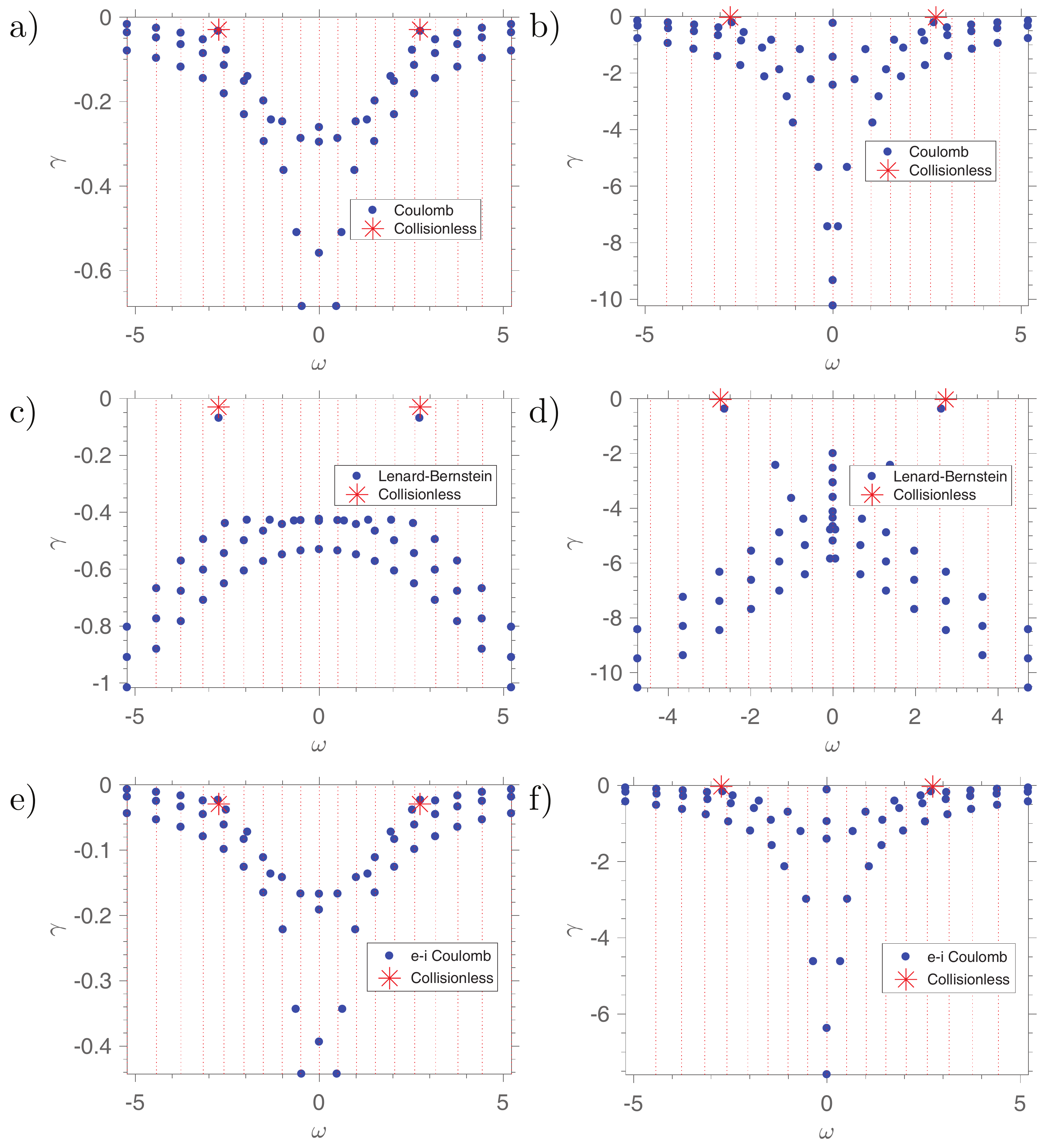}
    \caption{{Complete} eigenvalue spectrum of the truncated moment-hierarchy equation with $\alpha_D=0.09$ and $(P,J)=(18,2)$  {[yielding $(P+1)(J+1)=57$ eigenvalues]}, using the full-Coulomb collision operator (top), the Lenard-Bernstein operator (middle), and the electron-ion Coulomb operator only (bottom), with $\nu=0.1$ (left) and $\nu=1$ (right). The collisionless {least damped} solution is shown as a red marker, and the red vertical lines are the solutions of \cref{eq:collessol}.}
    \label{fig:coulroots}
\end{figure}

The resulting eigenvalue spectrum for the Coulomb collision case is shown in \cref{fig:coulroots} for $\nu=0.1$ (a) and $\nu=1$ (b), with $\alpha_D=0.09$ and $(P,J)=(18,2)$, together with the corresponding collisionless {Landau} root (red marker){, i.e., the least damped solution of \cref{eq:collisionless}}.
The resulting collisional spectrum is discrete, contrary to the continuous collisionless Van-Kampen spectrum, as noted in previous studies of weakly collisional plasma systems \citep{Ng1999,Bratanov2013}.
Figure \ref{fig:coulroots} shows that the damping rate of the Coulomb eigenmodes decreases with the corresponding frequency, which is possibly related to the fact that the collisional drag force decreases with the particle velocity in the Coulomb collision operator.
We also note that the least damped Coulomb {eigenvalue} in \cref{fig:coulroots} is not the one closest to the Landau collisionless solution, as there are modes with higher oscillation frequency $\omega$ that are less damped than the collisionless damping rate.
{These eigenvalue solutions, however, are related to eigenvectors that mainly involve moments $N^{pj}$ with large values of $p$ and $j$, and have therefore a negligible contribution to the {initial damping} of $N^{00}$ and $\phi$.}

Finally, the Coulomb eigenmode spectrum in \cref{fig:coulroots} includes modes with vanishing frequency and damping that increase{s} with $\nu$.
These modes correspond therefore to purely damped modes with a damping rate that at low collisionalities can be {comparable to} the collisionless Landau one.
These zero-frequency solutions have also been previously observed in the analysis of linear EPW when pitch-angle scattering effects are included \citep{Epperlein1992, Banks2016}, and correspond to the entropy mode studied in \cref{sec:timeevol}.

As an aside, we note that when the moment-hierarchy equation, \cref{eq:momenthierarchy}, is truncated at a higher $P$, i.e., using a higher number of Hermite polynomials, the number of eigenmodes with high frequency and small damping rate increases.
On the other hand, when the number of Laguerre polynomials, hence $J$, is increased, the eigenmode spectrum present and increasing number of modes with similar frequencies but increasingly higher damping rates.
However, as shown by \cref{fig:convstudy}, the damping rates $\gamma$ closest to the collisionless solution have negligible variation when $P$ and $J$ are increased (for $P\ge18$ and $J\ge0$ the variation is smaller than 3\%).

The eigenmode spectra using a Lenard-Bernstein collision operator are also shown in \cref{fig:coulroots} for $\nu=0.1$ (c) and $\nu=1$ (d), with $\alpha_D=0.09$ and $(P,J)=(18,2)$.
A clear difference is seen between the eigenmode spectra of the Coulomb and Lenard-Bernstein operators.
Contrary to the Coulomb case, the damping rate of the EPW modes increases with the frequency {$\omega$} when a Lenard-Bernstein collision operator is used.
Also, contrary to the Coulomb case, the Lenard-Bernstein root closest to the Landau collisionless root is the least damped one, as also noted in previous weakly-collisional studies of EPW \citep{Bratanov2013}.
%

Finally, the eigenvalue spectrum using the electron-ion Coulomb operator introduced in \cref{eq:cei} is shown in \cref{fig:coulroots} for $\nu=0.1$ (e) and $\nu=1$ (f).
The spectrum is {qualitatively} similar to the Coulomb one, with high frequency modes being less damped than modes with smaller oscillation frequency.
As for the Coulomb collision operator, such {frequency} dependence may be due to the dependence of the drag force on the particle velocity.
{Indeed}, the electron-ion collision operator contains a drag force that decreases with the particle velocity, similarly to the Coulomb operator.

We now estimate the frequency $\omega$ of the modes in \cref{fig:coulroots} with a damping rate $\gamma$ different than the ones closest to the collisionless roots, by noting that {the values of} $\omega$ {in \cref{fig:coulroots} are} seen to be weakly dependent on $\nu, \alpha_D$, and the collision operator for the range of values used.
We therefore solve the moment-hierarchy equation, \cref{eq:momenthierarchy}, in the $\phi=0$ limit, which effectively neglects the roots related to EPW.
{Furthermore, in order to retrieve purely oscillatory solutions, the collisional damping terms $C^{pj}$ in \cref{eq:momenthierarchy} are neglected.}
The time Fourier-transformed moment-hierarchy equation in the $\phi=C^{pj}=0$ limit reads
\begin{equation}
    \omega N^{pj} = \sqrt{\frac{p+1}{2}}N^{p+1 j}+\sqrt{\frac{p}{2}}N^{p-1 j}.
\label{eq:collessphilessmh}
\end{equation}
We recognize in \cref{eq:collessphilessmh} the recursion relation for the Hermite polynomials 
\begin{equation}
    N^{pj}=\frac{H_p(\omega)}{\sqrt{2^p p!}}.
\label{eq:solcollessphilessmh}
\end{equation}

The roots $\omega$ can be found by applying the truncation condition in \cref{eq:trunc} to the solution in \cref{eq:solcollessphilessmh}, yielding
\begin{equation}
    H_{P+1}(\omega) = 0.
\label{eq:collessol}
\end{equation}
The solutions $\omega$ in \cref{eq:collessol} are purely real, yielding frequencies that closely follow the ones observed in the eigenvalue spectra (red vertical lines in \cref{fig:coulroots}).

\begin{figure}
    \centering
    \includegraphics[width=.49\textwidth]{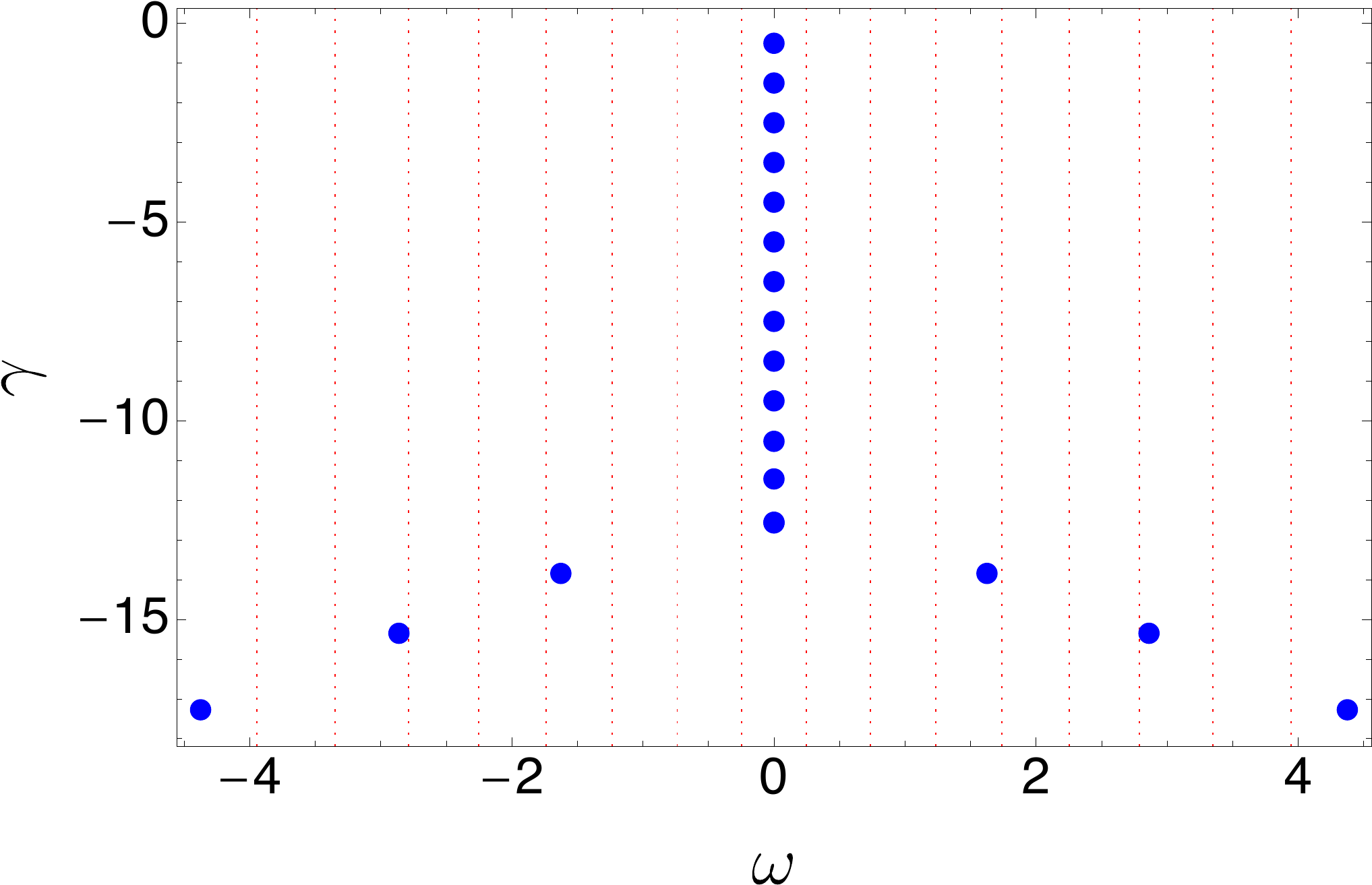}
    \includegraphics[width=.49\textwidth]{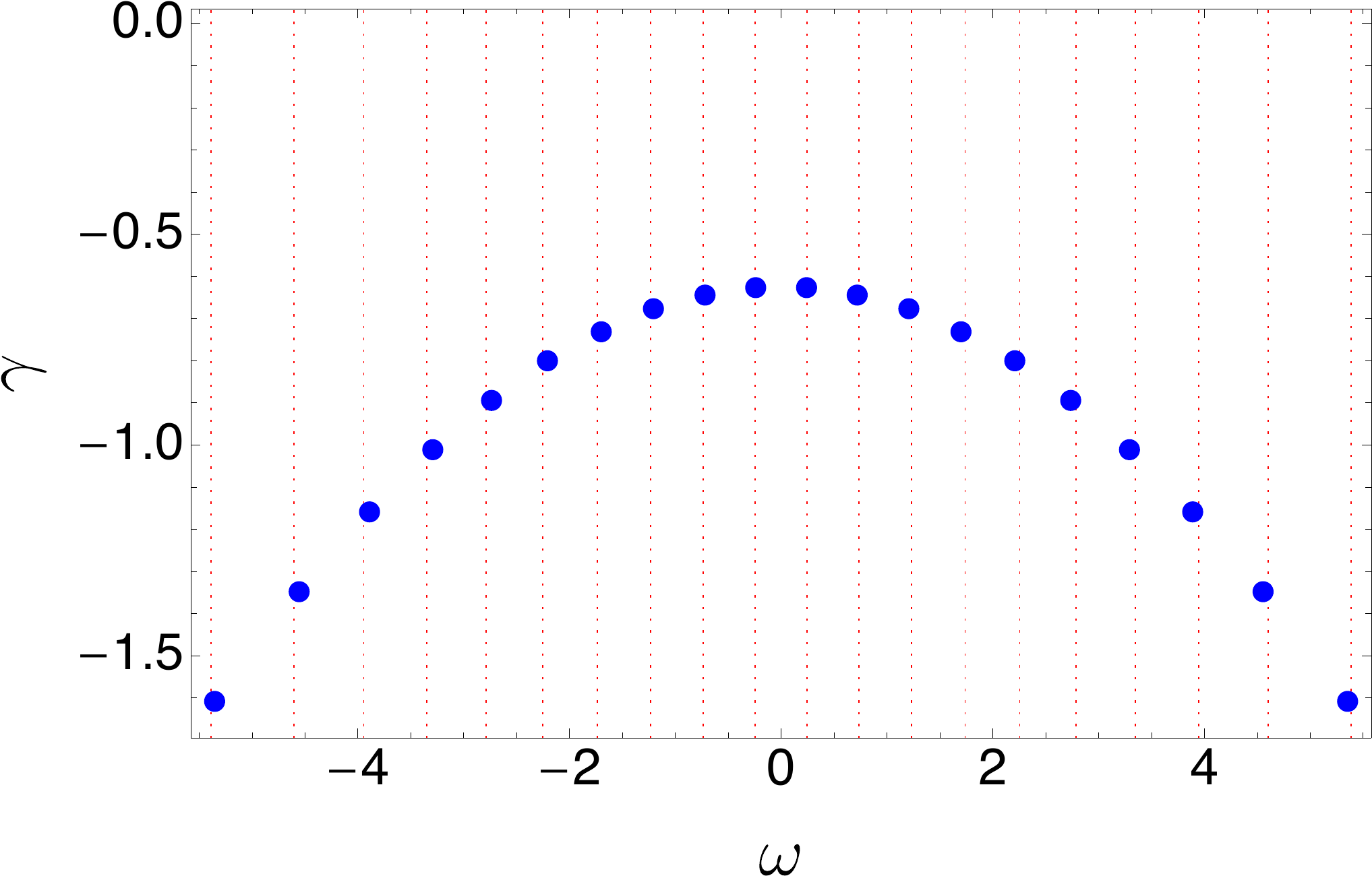}
    \caption{{Blue dots:} roots of the polynomial in \cref{eq:LBpolynomroot}, which corresponds to the solution of the Boltzmann equation with a Lenard-Bernstein collision operator where the distribution function is approximated by a truncated Hermite expansion, for $\lambda=50$ (left) and $\lambda=0.5$ (right) {(corresponding to $\nu=0.1$ and 1, respectively)} at $P=20$. {Red vertical lines: solutions of \cref{eq:collessol}.}}
    \label{fig:closedFormLBzeros}
\end{figure}

Finally, we present two tests to assess the validity of the results in \cref{fig:coulroots}, first for the Lenard-Bernstein case and then for the electron-ion case.
Focusing on the Lenard-Bernstein spectrum, we derive a polynomial in $\gamma$ whose roots closely follow the modes in  \cref{fig:coulroots} (c) and (d) that appear with damping rates larger than the ones of the two least damped roots.
Fourier transforming the Boltzmann equation, \cref{eq:boltzmanneq}, in time and in velocity-space similarly to \citet{Ng2004}, with $C(f)$ the Lenard-Bernstein collision operator, the following differential equation for $g(s)=\int_{-\infty}^{\infty} \exp(i s v_z- \gamma t+i\omega t) f_M \delta f_k dv_z dt$ is obtained
\begin{equation}
    g(s)\left(\gamma+i\omega+\frac{\nu}{2}s^2\right)+(1+ \nu s)\frac{d g(s)}{ds}=- s\frac{\sqrt{\pi}}{2 \alpha_D} e^{-\frac{s^2}{4}} g(0).
\label{eq:fouriervlb}
\end{equation}
We solve \cref{eq:fouriervlb} neglecting the coupling with the electrostatic potential $\phi$ by setting $\alpha_D \gg 1$ (or, equivalently, setting $\phi=0$ in the Boltzmann equation), and define $\lambda = \nu^{-2}/2$ and $\Gamma = \sqrt{2 \lambda}(\gamma+i\omega) -\lambda$, yielding
\begin{equation}
    g(s) = g(0)\left(1+\frac{s}{2\lambda}\right)^\Gamma e^{-\frac{s^2}{4}+s \lambda}.
\label{eq:fourvelsolLB}
\end{equation}
Similarly, Fourier transforming the Hermite-Laguerre expansion of $\left<\delta f_k\right>$ in velocity-space, we obtain
\begin{equation}
    g(s) = \sum_{p=0}^\infty \frac{i^p N^{p0}}{\sqrt{2^{p+1}p!}} s^p e^{-\frac{s^2}{4}}.
\label{eq:hermitevelfourier}
\end{equation}
Equating the two expressions above, we find
\begin{equation}
    N^{p0} = N^{00} (-i)^p  \sqrt{\frac{2^{p+1}}{p!}}\frac{d^p}{d s^p}\left[e^{s \lambda}\left(1+\frac{s}{2\lambda}\right)^\Gamma\right]_{s=0}.
\end{equation}
Therefore, the truncation condition in \cref{eq:trunc} in the $\phi=0$ limit is equivalent to imposing
\begin{equation}
    \frac{d^P}{d s^P}\left[e^{s \lambda}\left(1+\frac{s}{2\lambda}\right)^\Gamma\right]_{s=0}=0.
\label{eq:dercondition}
\end{equation}
Finally, we can rewrite \cref{eq:dercondition} as a polynomial in $\gamma+i\omega$
\begin{equation}
    \sum_{t=0}^P a_{Pt}(\lambda) (\gamma+i\omega)^t=0,
\label{eq:LBpolynomroot}
\end{equation}
with
\begin{equation}
    a_{pt}(\lambda) = \sum_{l=t}^p\sum_{n=l}^p \binom{p}{n}\binom{l}{t}s(n,l)\lambda^{p+l-n-t/2}(-1)^{l-t}2^{t/2},
\end{equation}
where $s(n,l)$ are the Stirling numbers of the first kind \citep{Moser1958,Qi2014}.
As shown in \cref{fig:closedFormLBzeros}, the polynomial expression in \cref{eq:LBpolynomroot} closely reproduces the eigenvalue spectrum observed in \cref{fig:coulroots} (c) and (d).
Therefore, although the coupling of the electron distribution function with $\phi$ is crucial to reproduce the EPW roots, additional modes in the eigenmode spectrum are related to solutions decoupled from the electrostatic potential $\phi$, subject to the truncation condition of the Hermite-Laguerre series, \cref{eq:trunc}, with frequencies similar to the ones of \cref{eq:collessol}.

\begin{figure}
    \centering
    \includegraphics[width=.49\textwidth]{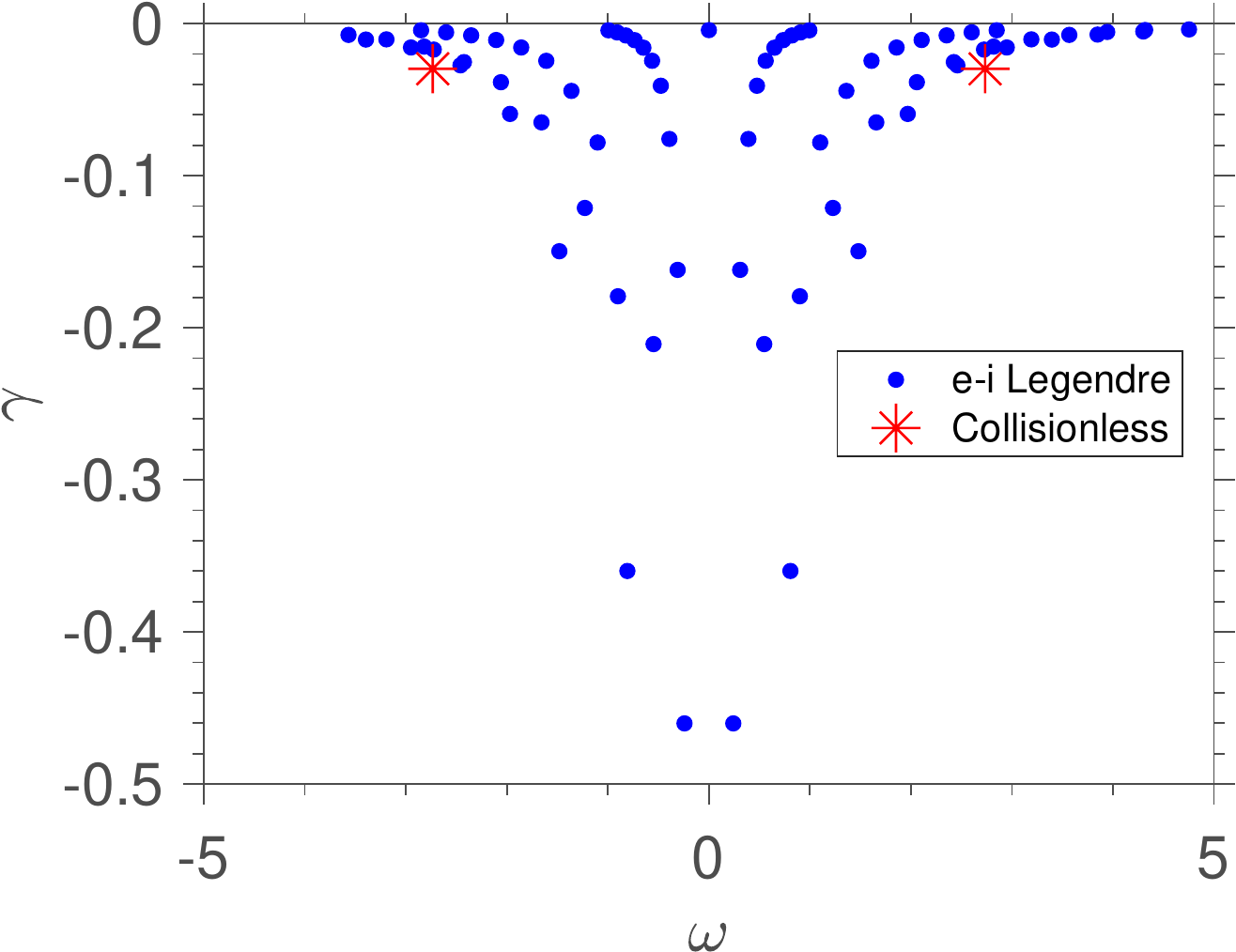}
    \includegraphics[width=.49\textwidth]{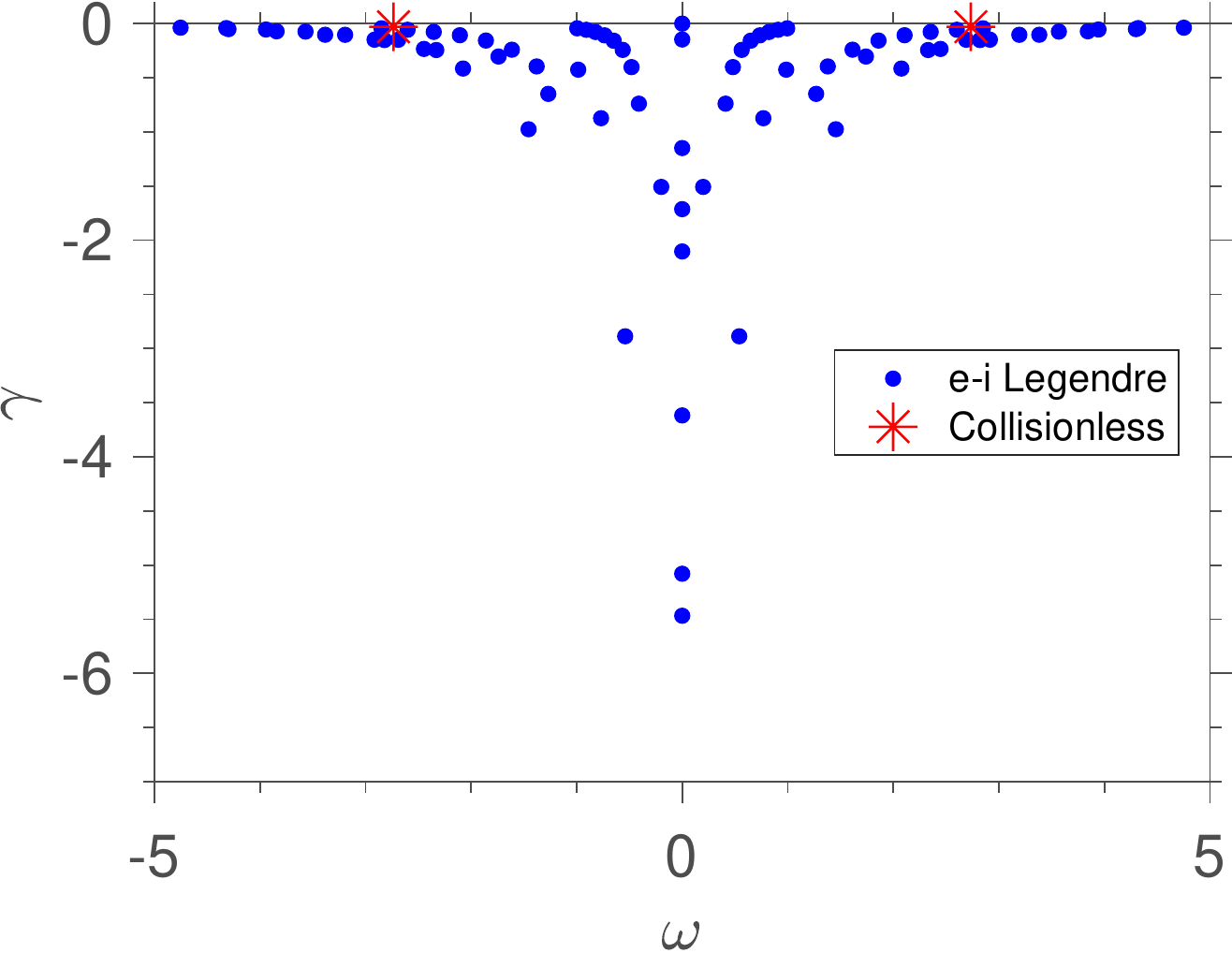}
    \caption{Eigenvalue spectrum of the truncated moment-hierarchy equation using an electron-ion Coulomb collision operator for $\nu=0.1$ (left) and $\nu=1.0$ (right) with $\alpha_D=0.09$, $n_v=12$ and $L=7$, with a Legendre decomposition. The collisionless solution is shown with a red marker.}
    \label{fig:eirootsLe}
\end{figure}

As a second test, to assess the validity of the eigenmode spectrum found with an electron-ion collision operator in \cref{fig:coulroots}, we solve the Boltzmann equation using a different set of basis functions, namely expanding $\left< \delta f_k \right>$ in Legendre polynomials, \cref{eq:legmomenthi}, and solving the resulting moment-hierarchy equation, \cref{eq:legmomenthi}, numerically.
In this case, the expansion of $\left<\delta f_k\right>$ in \cref{eq:legendreexp} is truncated at $l_{max}=L$ by setting $a_{L+1}=0$.
The velocity $v$ is discretized over an interval $[0,v_{\text{max}}]$ with an equidistant mesh {made of} $n_v$ points, and the integral estimated with a composite trapezoidal rule.
The resulting spectrum is shown in \cref{fig:eirootsLe}.
When compared with the Hermite-Laguerre spectrum in \cref{fig:coulroots} (e) and (f), the two spectra look qualitatively similar, confirming the validity of the Hermite-Laguerre approach.
However, a higher number of small-damped low-frequency solutions is observed when a Legendre decomposition is used.
The appearance of small-damped non-physical eigenmodes when using a finite-difference discretization in $v$ was also noted by \citet{Bratanov2013}, leading to the conclusion that, in general, a Hermite discretization of the distribution function is in fact superior to a finite difference one.
{Furthermore, for the values of $\nu=0.02$ and $\alpha_D=0.09$ where the Hermite-Laguerre formulation with $(P+1)(J+1)=19\times 3=57$ polynomials is seen to converge to the collisionless Landau solution in \cref{fig:coulLBcomp} with a relative difference of $\sim 16 \%$, when using a Legendre decomposition, a total of $n_v\times L \simeq 100$ equations is needed to yield a similar accuracy on $\gamma$.}

\section{Conclusion}
\label{sec:conclusions}

In this work, for the first time, the effect of full Coulomb collisions on electron-plasma waves is studied by taking into account both electron-electron and electron-ion collisions in {their} exact form.
The analysis is performed using an expansion of the distribution function and the Coulomb collision operator in a Hermite-Laguerre polynomial basis.
The proposed framework is particularly efficient, as the number of polynomials needed in order to obtain convergence is low enough to allow multiple scans to be performed, particularly a comparison between several collision operators at arbitrary collisionalities.
While the use of electron-ion collisions {alone} is seen to slightly decrease the damping rate with respect to full Coulomb collisions, the damping rate using a Lenard-Bernstein or a Dougherty collision operator is seen to yield deviations up to 50\% larger with respect to the Coulomb one.
An eigenmode analysis reveals major differences between the spectrum of full Coulomb and simplified collision operators.
The eigenspectrum reveals the presence of purely damped modes that, as shown, correspond to the entropy mode.
{In the collisional limit, the entropy mode is observed to set the long time behavior of the system with a damping rate smaller than the Landau damping one.}
We show that this mode needs a full-Coulomb collision operator for its proper description.
Furthermore, we find an analytical dispersion relation for the entropy mode that accurately reproduces the numerical results.

\section{Acknowledgements}

This work has been carried out within the framework of the EUROfusion Consortium and has received funding from the Euratom research and training programme 2014-2018 and 2019 - 2020 under grant agreement No 633053, and from Portuguese FCT - Fundação para a Ciência e Tecnologia, under grant PD/BD/105979/2014, carried out as part of the training in the framework of the Advanced Program in Plasma Science and Engineering (APPLAuSE, sponsored by FCT under grant No. PD/00505/2012) at Instituto Superior Técnico (IST).
S.G. acknowledges financial support from Fusenet.
N.F.L. was partially funded by US Department of Energy Grant no. DE-FG02-91ER54109.
This work was supported in part by the Swiss National Science Foundation.
The views and opinions expressed herein do not necessarily reflect those of the European Commission.

\bibliographystyle{jpp}
\bibliography{library}

\end{document}